\documentclass[12pt]{article}

\usepackage{amsmath}
\allowdisplaybreaks
\usepackage{amssymb}

\begin{document}

\baselineskip=17.0pt plus 0.2pt minus 0.1pt

\makeatletter
\@addtoreset{equation}{section}
\renewcommand{\theequation}{\thesection.\arabic{equation}}
\newcommand{\bm}[1]{\boldsymbol{#1}}
\newcommand{\calA}{\mathcal{A}}
\newcommand{\calB}{\mathcal{B}}
\newcommand{\calC}{\mathcal{C}}
\newcommand{\calE}{\mathcal{E}}
\newcommand{\calP}{\mathcal{P}}
\newcommand{\calM}{\mathcal{M}}
\newcommand{\calN}{\mathcal{N}}
\newcommand{\calV}{\mathcal{V}}
\newcommand{\calK}{\mathcal{K}}
\newcommand{\calF}{\mathcal{F}}
\newcommand{\calG}{\mathcal{G}}
\newcommand{\calH}{\mathcal{H}}
\newcommand{\calT}{\mathcal{T}}
\newcommand{\calU}{\mathcal{U}}
\newcommand{\calY}{\mathcal{Y}}
\newcommand{\calW}{\mathcal{W}}
\newcommand{\calL}{\mathcal{L}}
\newcommand{\calD}{\mathcal{D}}
\newcommand{\calO}{\mathcal{O}}
\newcommand{\calI}{\mathcal{I}}
\newcommand{\calQ}{\mathcal{Q}}
\newcommand{\calS}{\mathcal{S}}
\newcommand{\QB}{Q_\textrm{B}}
\newcommand{\nn}{\nonumber}
\newcommand{\drv}[2]{\frac{d #1}{d#2}}
\newcommand{\veps}{\varepsilon}
\newcommand{\eps}{\epsilon}
\newcommand{\ds}{\displaystyle}
\newcommand{\Ke}{K_{\veps}}
\newcommand{\invKe}{\frac{1}{\Ke}}
\newcommand{\Ue}{U_{\veps }}
\newcommand{\Ge}{G_{\veps }}
\newcommand{\Psie}{\Psi_{\veps}}
\newcommand{\calPe}{\calP_{\veps}}
\newcommand{\CR}[2]{\left[#1,#2\right]}
\newcommand{\ACR}[2]{\left\{#1,#2\right\}}
\newcommand{\Pmatrix}[1]{\begin{pmatrix} #1 \end{pmatrix}}
\newcommand{\tr}{\mathop{\rm tr}}
\newcommand{\Tr}{\mathop{\rm Tr}}
\newcommand{\p}{\partial}
\newcommand{\wh}[1]{\widehat{#1}}
\newcommand{\wt}[1]{\widetilde{#1}}
\newcommand{\ol}[1]{\overline{#1}}
\newcommand{\abs}[1]{\left| #1\right|}
\newcommand{\VEV}[1]{\left\langle #1\right\rangle}
\newcommand{\Drv}[2]{\frac{\p #1}{\p #2}}
\newcommand{\ket}[1]{| #1 \rangle}
\newcommand{\bra}[1]{\langle #1 |}
\newcommand{\braket}[2]{\langle #1 | #2 \rangle}
\newcommand{\KBc}{K\!Bc}
\newcommand{\fDrv}[2]{\frac{\delta #1}{\delta #2}}
\newcommand{\Qs}{\calQ_{\Psis}}
\newcommand{\Psis}{\Psi_{\!S}}
\newcommand{\Ngh}{N_\textrm{gh}}
\newcommand{\aw}{\alpha}
\newcommand{\eaK}{e^{-\aw K}}
\newcommand{\eaKe}{e^{-\aw\Ke}}
\newcommand{\dB}{\delta_\textrm{B}}
\newcommand{\whdB}{\wh{\delta}_\textrm{B}}
\newcommand{\uu}{\wh{u}}
\newcommand{\EOMe}{\textrm{EOM}_\veps}
\newcommand{\La}{L^{(a)}}
\newcommand{\Ra}{R^{(a)}}
\newcommand{\Lb}{L^{(b)}}
\newcommand{\Rb}{R^{(b)}}
\newcommand{\invveps}{\frac{1}{\veps}}
\newcommand{\cCR}[2]{\CR{#1}{#2}_{\textstyle{c}}}
\newcommand{\cR}[2]{\bigl[#1,#2\bigr]}
\newcommand{\Ser}{\mathop{\textrm{Ser}}}
\newcommand{\uoe}{\frac{u}{\veps}}
\newcommand{\eou}{(\veps/u)}
\newcommand{\PsiT}{\Psi_\textrm{T}}
\newcommand{\wA}{w_A}
\newcommand{\wB}{w_B}
\newcommand{\wC}{w_C}
\newcommand{\EA}{E_A}
\newcommand{\EB}{E_B}
\newcommand{\EC}{E_C}
\newcommand{\eveps}{e^{-\veps\left(\pi+\sum s\right)}}
\newcommand{\si}{\tilde{s}}
\newcommand{\sia}{\tilde{s}_a}
\newcommand{\sib}{\tilde{s}_b}
\newcommand{\sapib}{S_{a+\tilde{b}}}
\newcommand{\sbpia}{S_{b+\tilde{a}}}
\newcommand{\intfour}{\int\limits_{(s_a,s_b,\sia,\sib)}}

\makeatother
\begin{titlepage}

\title{
\hfill\parbox{3cm}{\normalsize KUNS-2593}\\[1cm]
{\Large\bf
BV Analysis of Tachyon Fluctuation around Multi-brane Solutions in
Cubic String Field Theory
}}

\author{
Hiroyuki {\sc Hata}\footnote{
{\tt hata@gauge.scphys.kyoto-u.ac.jp}}
\\[7mm]
{\it
Department of Physics, Kyoto University, Kyoto 606-8502, Japan
}
}

\date{{\normalsize November 2015}}
\maketitle

\begin{abstract}
\normalsize
We study whether the tachyon mode exists as a physical fluctuation
on the 2-brane solution and on the tachyon vacuum solution in cubic
open string field theory. Our analysis is based on the
Batalin-Vilkovisky formalism. We first construct a set of six
string states which corresponds to the set of fields and anti-fields
containing the tachyon field.
Whether the tachyon field can exist as a physical fluctuation is
determined by the $6\times 6$ matrix defining the anti-bracket in the
present sector. If the matrix is degenerate/non-degenerate, the
tachyon field is physical/unphysical. Calculations for the pure-gauge
type solutions in the framework of the $\KBc$ algebra and using the
$\Ke$-regularization lead to the expected results.
Namely, the matrix for the anti-bracket is degenerate/non-degenerate
in the case of the 2-brane/tachyon-vacuum solution.
Our analysis is not complete, in particular, in that we have not
identified the four-fold degeneracy of tachyon fluctuation on the
2-brane solution, and moreover that the present six states do not
satisfy the hermiticity condition.

\end{abstract}

\thispagestyle{empty}
\end{titlepage}

\section{Introduction}

After the discovery of exact tachyon vacuum solution \cite{Schnabl} in
cubic string field theory (CSFT) followed by its concise understanding
\cite{ES09} in terms of the $\KBc$ algebra \cite{Okawa}, there have
been considerable developments in the construction of multi-brane
solutions \cite{MS1,HKwn,MS2,HKinfty}.
The identification of a solution as the $n$-brane one representing $n$
pieces of D25-branes has been done from its energy density
consideration.
However, for the complete identification, we have to show that the
the physical excitations on the solution are those of the open string
and, in particular, that each excitation has $n^2$ degeneracy.
For the tachyon vacuum solution ($n=0$), a general proof has been
given for the absence of physical excitations \cite{EllSch}.
On the other hand, for $n$-brane solution with $n\ge 2$, no formal
existence proof nor an explicit construction of the excitations has
been given.\footnote{
See \cite{EM14}, for a construction of multi-brane
solutions and the fluctuation modes around them by introducing the
boundary condition changing operators.
}

In this paper, we present an explicit analysis of fluctuations around
multi-brane solutions in the framework of the Batalin-Vilkovisky (BV)
formalism \cite{BV1,BV2}.
Our analysis is not a complete one, but is rather a first step toward
the final understanding.
First, our analysis is restricted only to the tachyon vacuum solution
and the 2-brane one.
Second, we do not solve the general excitation modes on the
solution. Our analysis is restricted to the tachyon mode among all the
excitations.

Let us explain our analysis in more detail.
We are interested in the kinetic term of the action of CSFT expanded
around a multi-brane solution:
\begin{equation}
\calS_0=\frac12\int\!\Phi*\calQ\Phi ,
\label{eq:calS0_Intro}
\end{equation}
where $\calQ$ is the BRST operator in the background of the solution,
and $\Phi$ is the fluctuation around the solution.
Previous arguments have been mainly on the presence of the
homotopy operator $\calA$ on the tachyon vacuum solution satisfying
$\calQ\calA=\calI$ with $\calI$ being the identity string field. If
there exists a well-defined $\calA$, it implies that there are no
physical excitations at all.
In this paper, we carry out a different kind of analysis.
We consider a candidate tachyon field $\chi(x)$ as a fluctuation
around a class of multi-brane solutions, and examine whether $\chi$
represents a genuine physical excitation or it is unphysical.
In the former case, the lagrangian of $\chi$ contained in
\eqref{eq:calS0_Intro} should be the ordinary one:\footnote{
The space-time metric used in this paper is the mostly plus one;
$g_{\mu\nu}=\textrm{diag}(-1,1,1,\cdots,1)$.
}
\begin{equation}
\calL_\chi=-\frac12\left(\left(\p_\mu\chi\right)^2+m^2\chi^2\right) .
\label{eq:L_chi}
\end{equation}
On the other hand, if $\chi$ is unphysical, it should be a member of
unphysical BRST quartet fields $\left(\chi,C,\ol{C},\calB\right)$ with
the lagrangian given by a BRST-exact form \cite{KugoOjima}:
\begin{equation}
\calL_\textrm{quartet}=i\dB\!\left[
\ol{C}\left(\left(\Box-m^2\right)\chi-\frac12 \calB\right)\right]
=-\calB\left(\Box-m^2\right)\chi+\frac12 \calB^2
-i\ol{C}\left(\Box-m^2\right)C ,
\label{eq:L_quartet}
\end{equation}
where the BRST transformation $\dB$ (satisfying the nilpotency
$\dB^2=0$) is defined by
\begin{equation}
\dB\chi=C,\qquad \dB C=0,\qquad \dB\ol{C}=i\calB,\qquad \dB \calB=0 .
\end{equation}
In CSFT which has been constructed in the BV formalism, the
lagrangian for unphysical $\chi$ is not of the type
\eqref{eq:L_quartet} containing the auxiliary field $\calB$, but is
rather the one obtained by integrating out $\calB$:
\begin{equation}
\calL'_\textrm{quartet}=-\frac12\left[\left(\Box-m^2\right)\chi\right]^2
-i\ol{C}\left(\Box-m^2\right)C .
\label{eq:calL_q^prime}
\end{equation}
This is invariant under the redefined BRST transformation $\dB'$:
\begin{equation}
\dB'\chi=C,\qquad \dB' C=0,\qquad
\dB'\ol{C}=i\left(\Box-m^2\right)\chi .
\label{eq:dB^prime}
\end{equation}
Note that $\dB'$ is nilpotent only on-shell; in particular, we have
$(\dB')^2\ol{C}=i\left(\Box-m^2\right)C$.

Our analysis is carried out within the framework of the BV formalism.
We first construct six first-quantized string states $u_i(k)$ carrying
center-of-mass momentum $k_\mu$. These six states
correspond to the three fields $(\chi,C,\ol{C})$ in
\eqref{eq:calL_q^prime} as well as their anti-fields
$(\chi_\star,C_\star,\ol{C}_\star)$.
We call the six states $u_i$ the tachyon BV states.
Then, we examine the $6\times 6$ matrix $\omega_{ij}=\int\!u_i\,u_j$
given by the CSFT integration. In fact, this $\omega_{ij}$ is the
matrix defining the anti-bracket in the BV formalism, and it
determines whether $\chi$ is physical or unphysical.
If $\omega_{ij}$ is non-degenerate, namely, $\det\omega_{ij}\ne 0$,
$\chi$ is unphysical. More precisely, after the gauge-fixing by
removing the anti-fields, we obtain the lagrangian
\eqref{eq:calL_q^prime} of an unphysical system.
On the other hand, if $\omega_{ij}$ is degenerate, it implies that
the six states $u_i$ are not independent, and therefore, some of the
fields/anti-fields necessary for making $\chi$ unphysical are
missing. Concrete analysis shows that the lagrangian for $\chi$ in the
case of degenerate $\omega_{ij}$ is the physical one
\eqref{eq:L_chi}.

We consider multi-brane solutions in CSFT given formally as the
pure-gauge $U\QB U^{-1}$ with $U$ specified by a function $G(K)$
of $K$ in the $\KBc$-algebra (see \eqref{eq:UandU^-1}).
The point is that the eigenvalues of $K$ are in the range
$K\ge 0$, and various physical quantities associated with the solution
such as the energy density are not well-defined due to the
singularity at $K=0$. Therefore, we introduce the $\Ke$-regularization
of replacing $K$ in $U\QB U^{-1}$ by $\Ke=K+\veps$ with $\veps$ being
a positive infinitesimal \cite{HKwn}.
Then, the regularized solution $(U\QB U^{-1})_{K\to \Ke}$ is no longer
exactly pure-gauge, and the zero or the pole of $G(K)$ at $K=0$ is
interpreted as the origin of the non-trivial energy density
of the apparently pure-gauge configuration
\cite{MS1,HKwn,MS2,HKinfty}.\footnote{
The zero and pole of $G(K)$ at $K=\infty$ also make the pure-gauge
solutions non-trivial and more rich \cite{HKinfty}. However, we do not
consider this type of solutions in this paper.
}
Namely, $\veps$ from the infinitesimal violation of pure-gauge is
enhanced by $1/\veps$ from the singularity at $K=0$ to lead to
non-trivial results for the solution.

This phenomenon of $\veps\times(1/\veps)$ giving non-trivial results
also occurs in $\omega_{ij}$ in our BV analysis.
By the gauge transformation which transforms $U\QB U^{-1}$ to zero,
the regularized solution $(U\QB U^{-1})_{K\to \Ke}$ is transformed to
an apparently $O(\veps)$ quantity. Then, the corresponding BRST operator
$\calQ$ is almost equal to the original $\QB$; $\calQ=\QB+O(\veps)$.
Therefore, the matrix $\omega_{ij}$ for the six tachyon BV states is
reduced to a degenerate one if we simply put $\veps=0$ without taking
into account the singularity at $K=0$.
Namely, there exists a physical tachyon field on any $n$-brane
solution of the pure-gauge type in the naive analysis.
The total absence of physical excitations expected on the tachyon
vacuum should rather be a non-trivial phenomenon coming from
$\veps\times(1/\veps)\ne 0$.
Our interest here is whether this phenomenon does not occur on
$n$-branes with $n\ge 2$.

In CSFT, the meaning of the EOM, $\QB\Psi+\Psi^2=0$, is not so simple.
When we consider whether the EOM is satisfied by a candidate solution
$\Psis$, we have to specify the test string field $\PsiT$
and examine whether the EOM test,
$\int\!\PsiT*\left(\QB\Psis+\Psis^2\right)=0$, holds
or not. It is in general impossible that the EOM test holds for any
$\PsiT$, and the EOM test restricts both the solution and the
fluctuations around it.
For the pure-gauge type solutions mentioned above, the EOM
against itself (namely, $\PsiT=\Psis$) is satisfied only for
the tachyon vacuum solution and the 2-brane one (and, of course, for
the single-brane solution $\Psis=0$) \cite{HKwn}.
The correct value of the energy density can also be reproduced only
for these two solutions.
Therefore, in this paper, we carry out calculations of
$\omega_{ij}$ for these two kinds of solutions with $n=0$ and $2$.
Then, we need to take into account the EOM also in the construction of
the tachyon BV states $u_i$ on each solution.
For the BV analysis, the EOM must hold against the commutator
$\PsiT=\CR{u_i}{u_j}$ as we as $\PsiT=u_i$ themselves, and this is in
fact a non-trivial problem, in particular, for the 2-brane solution.
For devising such $u_i$, we multiply the naive expression of $u_0$
with the lowest ghost number by the functions of $\Ke$, $L(\Ke)$ and
$1/R(\Ke)$, from the left and the right, respectively, and define the
whole set of six $u_i$ by the operation of $\calQ$.
Then, we obtain the constraints on $L(\Ke)$ and $R(\Ke)$ from the
requirement of the EOM. The existence of $L(\Ke)$ and $R(\Ke)$
also affects the calculation of $\omega_{ij}$.

There is another important technical point in our BV analysis.
The matrix $\omega_{ij}=\int\!u_i u_j$ and the EOM test against the
commutator $\PsiT=\CR{u_i}{u_j}$ are functions of $k^2$ of the
momentum $k_\mu$ carried by $u_i$. Then, a problem arises:
Some of these quantities contain terms depending on $\veps$ of the
$\Ke$-regularization in a manner such as $\veps^{\min(2k^2-1,1)}$,
which diverges in the limit $\veps\to 0$ for a smaller $k^2$
and tends to zero for a larger $k^2$.
Therefore, we {\em define} them as the ``analytic continuation'' from
the region of sufficiently large $k^2$ (namely, sufficiently
space-like $k_\mu$) to drop this type of $\veps$-dependent terms.

Next, we comment on the ``cohomology approach'' to the problem of
physical fluctuation around a multi-brane solution.
In this approach, we consider the BRST cohomology
$\mathrm{Ker}\calQ/\mathrm{Im}\calQ$, namely, 
we solve $\calQ u_1(k)=0$ for $u_1(k)$ which carries ghost number one
and is not $\calQ$-exact.
However, the meaning of (non-)equality in $\calQ u_1=0$ and
$u_1\ne\calQ(*)$ is subtle for multi-brane solutions of the pure-gauge
type discussed in this paper due to the singularity at $K=0$. To make
these equations precise, we should introduce the $\Ke$-regularization
and consider their inner-products (CSFT integrations) with states in
the space of fluctuations.
We would like to stress that our BV analysis indeed gives
information for solving the BRST cohomology problem within the
$\Ke$-regularization. (The present BV analysis can identify some
of the non-trivial elements of $\mathrm{Ker}\calQ/\mathrm{Im}\calQ$.
However, it cannot give the complete answer to the cohomology problem
since we consider only a set of trial BV states.)
We will explain the interpretation of our results of the BV analysis
in the context of the cohomology approach in Secs.\
\ref{sec:action_2b} and \ref{sec:omega_tv}.
We also comment that the analysis of the BRST cohomology around the
tachyon vacuum by evaluating the kinetic term of the action of the
fluctuation in the level truncation approximation
\cite{HataTera,EllwoodTaylor,GiustoImbimbo} has some relevance to
the present BV approach.

Then, finally in the Introduction, we state our results obtained in
this paper. For the tachyon vacuum solution, we find that the matrix
$\omega_{ij}$ is non-degenerate. This implies that our candidate
tachyon field is an unphysical one belonging to a BRST quartet.
On the other hand, for the 2-brane solution, $\omega_{ij}$ turns out
to be degenerate, implying that the tachyon field is a physical one.
These results are both what we expect for each solution.
However, we have not succeeded in identifying the whole of the $2^2$
tachyon fields which should exist on the 2-brane solution.
In addition, the six tachyon BV states in this paper have a problem
that they do not satisfy the hermiticity requirement
(see Sec.\ \ref{sec:hermiticity}).

The organization of the rest of this paper is as follows.
In Sec.\ 2, we recapitulate the BV formalism used in this paper,
and give examples of the BV states on the unstable vacuum.
In Sec.\ 3, we present the construction of the six tachyon BV states
on a generic pure-gauge type solution, and prepare various formulas
necessary for the BV analysis.
In Sec.\ 4, we carry out the calculation of the EOM against $u_i$ and
$\CR{u_i}{u_j}$ and of each component of $\omega_{ij}$ on the 2-brane
solution to confirm the existence of a physical tachyon field. Next,
in Sec.\ 5, we repeat the same analysis for the tachyon vacuum
solution. There we find that the candidate tachyon field is
unphysical. We summarize the paper and discuss future problems in
Sec.\ 6. In the Appendices, we present various technical details used
in the text.

\section{BV formalism for CSFT}

The action of CSFT on the unstable vacuum \cite{Witten},\footnote{
We have put the open string coupling constant equal to one.
}
\begin{equation}
S[\Psi]=\int\!\left(\frac12\,\Psi*\QB\Psi+\frac13\,\Psi^3\right) ,
\label{eq:S}
\end{equation}
satisfies the BV equation:
\begin{equation}
\int\!\left(\fDrv{S}{\Psi}\right)^2=0 .
\label{eq:BVeq_S}
\end{equation}
Concretely, we have
\begin{equation}
\fDrv{S}{\Psi}=\QB\Psi+\Psi^2 ,
\label{eq:fDrv_S_Psi}
\end{equation}
and the BV equation holds due to (i) the nilpotency $\QB^2=0$
of the BRST operator $\QB$, (ii) the derivation property of $\QB$
on the $*$-product, (iii) the property $\int\!\QB(\cdots)=0$,
(iv) the associativity of the $*$-product, 
and (v) the cyclicity
$\int\!A_1*A_2=(-1)^{A_1A_2}\int\!A_2*A_1$ valid for any two string
fields $A_1$ and $A_2$.\footnote{
$(-1)^{A}=+1$ ($-1$) when $A$ is Grassmann-even (-odd).
}
The BV equation is a basic requirement in the construction
of gauge theories including SFT. The BV equation implies the gauge
invariance of the action. Moreover, it gives a consistent way of
gauge-fixing and quantization of the theory.

In this paper, we are interested in CSFT expanded around a non-trivial
solution $\Psis$ satisfying the EOM:
\begin{equation}
\QB\Psis+\Psis^2=0 .
\label{eq:EOM}
\end{equation}
Expressing the original string field $\Psi$ in \eqref{eq:S} as
\begin{equation}
\Psi=\Psis+\Phi ,
\label{eq:Psi=Psis+Phi}
\end{equation}
with $\Phi$ being the fluctuation, we obtain
\begin{equation}
S[\Psi]=S[\Psis]+\int\!\Phi*\left(\QB\Psis+\Psis^2\right)
+\calS_{\Psis}[\Phi] .
\label{eq:S=Ssol+calS}
\end{equation}
The second term on the RHS of \eqref{eq:S=Ssol+calS} should vanish due
to the EOM \eqref{eq:EOM}. However, for multi-brane solutions in CSFT,
this EOM term cannot vanish for all kinds of fluctuations $\Phi$ as
stated in the Introduction. This is the case even for the tachyon
vacuum solution. In this paper, we restrict the fluctuation $\Phi$
around $\Psis$ to those for which the EOM term of
\eqref{eq:S=Ssol+calS} vanishes. We will see later that the EOM term
must also vanish against the commutator among the fluctuations.

The last term of \eqref{eq:S=Ssol+calS} is the action of the
fluctuation:
\begin{equation}
\calS_{\Psis}[\Phi]
=\int\!\left(\frac12\,\Phi*\Qs\Phi+\frac13\,\Phi^3\right) .
\label{eq:calSs}
\end{equation}
The only difference between the two actions \eqref{eq:S} and
\eqref{eq:calSs} is that the BRST operator $\QB$ in the former is
replaced with $\Qs$, the BRST operator around the solution $\Psis$.
The operation of $\Qs$ on any string field $A$ with a generic ghost
number is defined by
\begin{equation}
\Qs A=\QB A+\Psis*A-(-1)^A A*\Psis .
\label{eq:QsA}
\end{equation}
The BV equation for $\calS_{\Psis}$,
\begin{equation}
\int\!\left(\fDrv{\calS_{\Psis}}{\Phi}\right)^2=0 ,
\label{eq:BVeq_calS}
\end{equation}
which is formally equivalent to \eqref{eq:BVeq_S} for the original
$S$, also holds since $\Qs$ satisfies the same three basic properties
as $\QB$ does; (i), (ii) and (iii) mentioned below
\eqref{eq:fDrv_S_Psi}.
Among them, the nilpotency $\Qs^2=0$ is a consequence of
the EOM; namely, we have from \eqref{eq:QsA}
\begin{equation}
\Qs^2 A=\CR{\QB\Psis+\Psis^2}{A} .
\label{eq:Qs^2A=}
\end{equation}
On the other hand, the other two properties (ii) and (iii) hold for
any $\Psis$ irrespectively of whether it satisfies the EOM or not.
In the following, we omit the subscript $\Psis$ in $\calS_{\Psis}$
unless necessary.

\subsection{BV equation in terms of component fields}

Here, we consider the BV equation \eqref{eq:BVeq_calS} for the action
\eqref{eq:calSs} in terms of the component fields.\footnote{
See, for example, \cite{Schwarz,HataZwiebach} for the BV formalism for
a general supermanifold of fields and anti-fields.
The matrix $\omega_{ij}$ in \cite{HataZwiebach} corresponds to
$(-1)^{\varphi^i}\omega_{ij}$ in this paper.
}
Let $\{u_i(k)\}$ be a ``complete set'' of states of fluctuation around
$\Psis$ (here, we take as $\Psis$ a translationally invariant
solution, and $k_\mu$ is the center-of-mass momentum of the
fluctuation).
Note that each $u_i(k)$ is a string field.
Then, we expand the fluctuation field $\Phi$ as
\begin{equation}
\Phi=\int_k\sum_i u_i(k)\,\varphi^i(k) ,
\label{eq:Phi=sumu_ivarphi^i}
\end{equation}
where $\varphi^i(k)$ is the component field corresponding to the state
$u_i(k)$, and $\int_k$ is short for $\int\!d^{26}k/(2\pi)^{26}$.
In \eqref{eq:Phi=sumu_ivarphi^i}, $u_i(k)$ may carry any ghost number
$\Ngh(u_i)$, and the ghost number of the corresponding $\varphi^i$
must satisfy
\begin{equation}
\Ngh(u_i)+\Ngh(\varphi^i)=\Ngh(\Phi)=1 .
\label{eq:Nghu+Nghvarphi=1}
\end{equation}
Then, we define the matrix $\omega_{ij}(k)$ and its inverse
$\omega^{ij}(k)$ by
\begin{equation}
\int\!u_i(k')\,u_j(k)=\omega_{ij}(k)\times
(2\pi)^{26}\delta^{26}(k'+k) ,
\label{eq:omega}
\end{equation}
and
\begin{equation}
\sum_j\omega^{ij}(k)\,\omega_{jk}(k)=\delta^i{}_k .
\end{equation}
Here, we are assuming that $\omega_{ij}$ is {\em non-degenerate},
namely, that the inverse matrix $\omega^{ij}$ exists.\footnote{
\label{fn:non-degenerate}
Precisely speaking, our assumption here is that $\det\omega_{ij}(k)$
is not identically equal to zero as a function of $k_\mu$.
$\omega_{ij}$ being degenerate at some points in the $k_\mu$ space is
allowed.
}
In particular, the number of the basis $u_i(k)$ must be even.
Note that $\omega_{ij}$ and $\omega^{ij}$ are non-vanishing only for
$(i,j)$ satisfying $\Ngh(u_i)+\Ngh(u_j)=3$, and therefore,
\begin{equation}
\Ngh(\varphi^i)+\Ngh(\varphi^j)=-1 .
\end{equation}
Note also that these matrices are symmetric in the following sense:
\begin{equation}
\omega_{ij}(k)=\omega_{ji}(-k),
\qquad
\omega^{ij}(k)=\omega^{ji}(-k) .
\end{equation}
The completeness relation of the set $\{u_i\}$ reads
\begin{equation}
A=\int_k\sum_{i,j} u_i(k)\,\omega^{ij}(k)\int\!u_j(-k)*A ,
\label{eq:completeness}
\end{equation}
for any string field $A$, and hence we have\footnote{
The sign factor $(-1)^{\varphi^j}$ in \eqref{eq:fDrvcalS_Phi} is due
to the fact that the CSFT integration $\int$ is Grassmann-odd.
In this paper, $\delta/\delta \varphi^j$ for a Grassmann-odd
$\varphi^j$ is defined to be the left-derivative
$\overrightarrow{\delta}/\delta \varphi^j$.
}
\begin{equation}
\fDrv{}{\Phi}=\int_k\sum_{i,j}u_i(k)\,
\omega^{ij}(k)\,(-1)^{\varphi^j}
\fDrv{}{\varphi^j(-k)} .
\label{eq:fDrvcalS_Phi}
\end{equation}
Using \eqref{eq:fDrvcalS_Phi} in \eqref{eq:BVeq_calS}, we obtain the
BV equation in terms of the component fields:
\begin{equation}
\int_k\sum_{i,j}\omega^{ij}(k)\,\fDrv{\calS}{\varphi^i(k)}
\,\fDrv{\calS}{\varphi^j(-k)}=0 .
\label{eq:BV_varphi}
\end{equation}

It is convenient to take the Darboux basis where the matrix
$\omega_{ij}(k)$ takes the following form:
\begin{equation}
\omega_{ij}(k)=\Pmatrix{0 & D(-k) \\ D(k) & 0},
\qquad
D(k)=\textrm{diag}\left(a^1(k),a^2(k),\cdots\right) .
\label{eq:omega_Darboux}
\end{equation}
Denoting the corresponding component fields, namely, the pair of
fields and anti-fields, as $\{\phi^i(k),\phi^i_\star(k)\}$ with the
index $i$ running only half of that for $\{\varphi^i\}$, the BV
equation \eqref{eq:BV_varphi} reads
\begin{equation}
\int_k\sum_{i}a^i(k)^{-1}\,\fDrv{\calS}{\phi^i_\star(-k)}
\,\fDrv{\calS}{\phi^i(k)}=0 .
\label{eq:BV_Darboux}
\end{equation}
Then, the gauge-fixed action $\wh{\calS}$ and the BRST transformation
$\whdB$ under which $\wh{\calS}$ is invariant are given by
\begin{equation}
\wh{\calS}[\phi]=\calS\bigr|_L ,
\qquad
\whdB \phi^i=i\left.
a^i(k)^{-1}\,\fDrv{\calS}{\phi^i_\star(-k)}\right|_L ,
\label{eq:whSandwhdB}
\end{equation}
where $|_L$ denotes the restriction to the Lagrangian submanifold
defined by the gauge-fermion $\Upsilon[\phi]$:
\begin{equation}
L:\quad\phi^i_\star=\fDrv{\Upsilon[\phi]}{\phi^i} .
\end{equation}
The simplest choice for $\Upsilon$ is of course $\Upsilon=0$.

\subsection{Examples of BV basis on the unstable vacuum}

For CSFT on the unstable vacuum, the BV basis $\{u_i(k)\}$ consists of
an infinite number of first quantized string states of all ghost
numbers. Though the whole BV basis is infinite dimensional, we can
consider a subbasis with non-degenerate $\omega_{ij}$ and consisting
of a finite number of states which are connected by the operation of
$\QB$ and are orthogonal (in the sense of $\omega_{ij}=0$) to any
states outside the subbasis.

Here, we present two examples of BV subbasis with non-degenerate
$\omega_{ij}$.
For our later purpose, we present them using the $\KBc$ algebra in the
sliver frame. The $\KBc$ algebra and the correlators in the sliver
frame are summarized in Appendix \ref{app:KBc}.
In the rest of this paper, we omit ``sub'' for the BV subbasis and
simply write ``BV basis'' since we will not consider the full BV
basis.

\subsubsection{Unphysical BV basis of photon longitudinal mode}
Our first example is the unphysical BV basis associated with
the longitudinal mode of the photon on the unstable vacuum
$\Psis=0$. Namely, we consider the unphysical model obtained
by restricting the photon field to the pure-gauge,
$A_\mu(x)=\p_\mu\chi(x)$.
The corresponding BV basis consists of the following six states:
\begin{alignat}{3}
\Ngh&=0:&\ u_0(k)&=\frac{1}{\sqrt{2}}\eaK\,V_k\,\eaK ,
\nn\\
\Ngh&=1:&\ u_{1A}(k)&=\frac{i}{\sqrt{2}}\,\eaK c\CR{K}{V_k}\eaK,
&u_{1B}(k)&=\frac{-i}{\sqrt{2}}\,\eaK\CR{K}{c}V_k\,\eaK ,
\nn\\
\Ngh&=2:&\ u_{2A}(k)
&=\frac{i}{\sqrt{2}}\,\eaK c\CR{K}{c}\CR{K}{V_k}\eaK,
&u_{2B}(k)&=\frac{i}{\sqrt{2}}\,\eaK c\CR{K}{\CR{K}{c}}V_k\,\eaK ,
\nn\\
\Ngh&=3:&\ u_3(k)&=\frac{1}{\sqrt{2}}
\,\eaK c\CR{K}{\CR{K}{c}}\CR{K}{c}V_k\,\eaK ,
\label{eq:sixstates_long}
\end{alignat}
where $\aw$ is a constant,
\begin{equation}
\aw=\frac{\pi}{4} ,
\label{eq:a}
\end{equation}
and $V_k$ is the vertex operator
of momentum $k_\mu$ at the origin:
\begin{equation}
V_k=e^{ik_\mu X^\mu(0,0)} .
\label{eq:V_k}
\end{equation}
These six states $u_i(k)$ are all chosen to be hermitian in the sense
that
\begin{equation}
u_i(k)^\dagger=u_i(-k) .
\label{eq:u^dagger=u}
\end{equation}
Among the six $u_i$, $u_{1A}$ is the photon state with longitudinal
polarization $k_\mu$.
The operation of $\QB$ on the six states \eqref{eq:sixstates_long} is
given as follows:
\begin{align}
i\QB u_0(k)&=u_{1A}(k)-k^2 u_{1B}(k) ,
\nn\\[2mm]
\QB\!\Pmatrix{u_{1A}\\ u_{1B}}&=-\Pmatrix{k^2 \\ 1}
\bigl(u_{2A}(k)+u_{2B}(k)\bigr) ,
\nn\\[2mm]
i\QB\!\Pmatrix{u_{2A}(k)\\ u_{2B}(k)}&=\Pmatrix{1\\ -1}k^2 u_3(k) ,
\nn\\[2mm]
\QB u_3(k)&=0 .
\label{eq:QBu_long}
\end{align}
The non-trivial components of the $6\times 6$ matrix $\omega_{ij}(k)$
are given by 
\begin{equation}
\omega_{0,3}=-1,
\qquad
\Pmatrix{\omega_{1A,2A} & \omega_{1A,2B}\\
\omega_{1B,2A} & \omega_{1B,2B}}
=\Pmatrix{k^2 & 0 \\ 0 & 1} ,
\label{eq:omega_long}
\end{equation}
and therefore $\omega_{ij}$ is non-degenerate.\footnote{
Though $\omega_{ij}(k)$ is degenerate at $k^2=0$, this is not a
problem as we mentioned in footnote \ref{fn:non-degenerate}.
}
Moreover, the present basis $\{u_i\}$ is already Darboux as seen
from \eqref{eq:omega_long}.
Then, expanding the string field $\Psi$ as
\begin{equation}
\Psi=\int_k\!\Bigl\{
u_0(k)\,C(k)+u_{1A}(k)\,\chi(k)+u_{1B}(k)\,\ol{C}_\star(k)
+u_{2A}(k)\,\chi_\star(k)+u_{2B}(k)\,\ol{C}(k)+u_3(k)\,C_\star(k)
\Bigr\} ,
\end{equation}
and using \eqref{eq:QBu_long} and \eqref{eq:omega_long}, we find that
the kinetic term of the CSFT action \eqref{eq:S} is given by
\begin{align}
&S_0=\frac12\int\!\Psi*\QB\Psi
\nn\\
&=\int_k\left\{
-\frac12\left(k^2\chi(-k)+\ol{C}_\star(-k)\right)
\left(k^2\chi(k)+\ol{C}_\star(k)\right)
+ik^2\left(\ol{C}(-k)-\chi_\star(-k)\right)C(k)
\right\} .
\label{eq:S0_long}
\end{align}
Finally, the gauge-fixed action $\wh{S}_0$ and the BRST transformation
$\whdB$ in the gauge $\ol{C}_\star=\chi_\star=C_\star=0$ are given
using \eqref{eq:whSandwhdB} by
\begin{equation}
\wh{S}_0=\int_k\left\{-\frac12\,k^2\chi(-k)\,k^2\chi(k)
+ik^2\ol{C}(-k) C(k)\right\} ,
\end{equation}
and
\begin{equation}
\whdB\chi(k)=C(k),
\qquad
\whdB\ol{C}(k)=-ik^2\chi(k),
\qquad
\whdB C(k)=0 .
\end{equation}
This is the $m^2=0$ version of the unphysical system given in
\eqref{eq:calL_q^prime} and \eqref{eq:dB^prime}.

\subsubsection{BV basis of the tachyon mode}

Our second example is the BV basis for the tachyon mode on the
unstable vacuum. It consists only of two states: the tachyon state
$u_1$ and its BRST-transform $u_2$:
\begin{equation}
\Ngh=1:\ u_1(k)=\eaK c\,V_k\,\eaK,
\qquad
\Ngh=2:\ u_2(k)=\eaK cKc\,V_k\,\eaK ,
\end{equation}
with
\begin{equation}
\QB\,u_1(k)=-\left(k^2-1\right)u_2(k),
\qquad
\QB\,u_2(k)=0 .
\end{equation}
The $2\times 2$ matrix $\omega_{ij}$ is non-degenerate since we have
\begin{equation}
\omega_{1,2}(k)=1 .
\end{equation}
Expressing the string field as
\begin{equation}
\Psi=\int_k\bigl(u_1(k)\phi(k)+u_2(k)\phi_\star(k)\bigr) ,
\label{eq:Psi=sum_u_phi_tach_pert}
\end{equation}
the kinetic term reads
\begin{equation}
S_0=-\frac12\int_k\phi(-k)\left(k^2-1\right)\phi(k) ,
\label{eq:S0_tach_pert}
\end{equation}
which does not contain the anti-field $\phi_\star$.
The gauge-fixed action $\wh{S}_0$ is the same as $S_0$, the ordinary
kinetic term of the tachyon field $\phi$. The BRST transformation of
$\phi$ is of course equal to zero;
$\whdB\phi=i\,\delta S_0/\delta\phi_\star|_L=0$.

\section{Tachyon BV states around a multi-brane solution}
\label{sec:BVbasis}

We consider the fluctuation around a multi-brane solution
$\Psie$ given as the $\Ke$-regularization of the pure-gauge
$U\QB U^{-1}$ \cite{HKwn}:
\begin{equation}
\Psie=\left(U\QB U^{-1}\right)_\veps
=c\frac{\Ke}{\Ge}Bc\left(1-\Ge\right) ,
\label{eq:Psie}
\end{equation}
where $U$ and its inverse $U^{-1}$ are specified by a function $G(K)$
of $K$:
\begin{equation}
U=1-Bc\left(1-G(K)\right),
\qquad
U^{-1}=1+\frac{1}{G(K)}Bc\left(1-G(K)\right) .
\label{eq:UandU^-1}
\end{equation}
Here and in the following, $\calO_\veps$ for a quantity $\calO$
containing $K$ denotes the $\Ke$-regularized one;
$\calO_\veps=\calO\bigr|_{K\to \Ke=K+\veps}$.
Therefore, we have $\Ge\equiv G(\Ke)$ in \eqref{eq:Psie}.
Although the EOM is satisfied automatically by the pure-gauge
$U\QB U^{-1}$, the $\Ke$-regularization breaks the EOM by the
$O(\veps)$ term:
\begin{equation}
\QB\Psie+\Psie*\Psie=\veps\times
c\frac{\Ke}{\Ge}c\left(1-\Ge\right) .
\label{eq:EOM_Psie}
\end{equation}
As we saw in \cite{HKwn}, this $O(\veps)$ breaking of the
EOM can be enhanced by the singularity at $K=0$ to lead to non-trivial
results for the EOM against $\Psie$ itself:
\begin{equation}
\int\!\Psie*\left(\QB\Psie+\Psie^2\right)
=\veps\times\int\!Bc\Ge c\frac{\Ke}{\Ge}c\Ge c\frac{\Ke}{\Ge} .
\label{eq:Psie*EOM}
\end{equation}
We found that \eqref{eq:Psie*EOM} vanishes for $G(K)$ having a simple
zero, a simple pole or none at all at $K=0$, which we expect to
represent the tachyon vacuum, the 2-brane and the 1-brane,
respectively,
from their energy density values.
For $G(K)$ with higher order zero or
pole at $K=0$, \eqref{eq:Psie*EOM} becomes non-vanishing.
Therefore, in this paper, we consider the following two $G(K)$ as
concrete examples:
\begin{align}
G_\textrm{tv}(K)=\frac{K}{1+K},
\qquad
G_\textrm{2b}(K)=\frac{1+K}{K} ,
\label{eq:twoG}
\end{align}
which correspond to the tachyon vacuum and the 2-brane,
respectively.

For our purpose of studying the fluctuation, it is more convenient to
gauge-transform $\Psie$ by
$\Ue^{-1}\left[=\left(\Ue\right)^{-1}=\left(U^{-1}\right)_\veps\right]$
to consider
\begin{equation}
\calPe=\Ue^{-1}\left(\Psie+\QB\right)\Ue
=\Ue^{-1}\left(\Psie-\Ue\QB\Ue^{-1}\right)\Ue
=\veps\times\frac{1}{\Ge}c\,\Ge Bc\left(1-\Ge\right) .
\label{eq:calPe}
\end{equation}
Note that $\calPe$ is apparently of $O(\veps)$ since, without the
$\Ke$-regularization, the present gauge transformation transforms the
pure-gauge $U\QB U^{-1}$ back to zero.
The fluctuation around $\calPe$ and that around $\Psie$ are
related by
\begin{equation}
\calS_{\calPe}[\Phi]=\calS_{\Psie}[\Ue\Phi\Ue^{-1}] ,
\end{equation}
for $\calS_{\Psis}$ of \eqref{eq:calSs}. Note the following property
of $\Qs$ \eqref{eq:QsA}:
\begin{equation}
\calQ_{V^{-1}(\Psis+\QB)V}(V^{-1}AV)=V^{-1}\left(\Qs A\right)V .
\end{equation}
The EOM of $\calPe$ is given by
\begin{equation}
\QB\calPe+\calPe^2=
\Ue^{-1}\left(\QB\Psie+\Psie^2\right)\Ue
=\veps\times\frac{1}{\Ge}c\left(\Ke c\Ge-\Ge c\Ke\right)
Bc\left(1-\Ge\right) .
\label{eq:EOM_calPe}
\end{equation}
Though the EOM against the solution itself, \eqref{eq:Psie*EOM}, is
not a gauge-invariant quantity, we have confirmed that
$\int\!\calPe*\left(\QB\calPe+\calPe^2\right)$ vanishes in the limit
$\veps\to 0$ for the two $G(K)$ in \eqref{eq:twoG}.

\subsection{Six tachyon BV states around $\calPe$}
\label{sec:six_tachBVstates}

We are interested in whether physical fluctuations exist or not around
the classical solutions $\calPe$ specified by two $G(K)$ in
\eqref{eq:twoG}. Our expectation is of course that there
are no physical fluctuations at all for $G_\textrm{tv}$, while
there are quadruplicate of physical fluctuations for
$G_\textrm{2b}$.
In this paper, we consider this problem in the framework of
the BV formalism by focusing on the tachyon mode.
In the following, $\calQ$ denotes $\calQ_{\calPe}$,
the BRST operator around $\calPe$:
\begin{equation}
\calQ A=\calQ_{\calPe}A=\QB A+\calPe*A-(-1)^A A*\calPe .
\label{eq:calQ}
\end{equation}
Accordingly, $\calS[\Phi]$ denotes $\calS_{\calPe}[\Phi]$,
the action of the fluctuation $\Phi$ around $\calPe$:
\begin{equation}
\calS[\Phi]=\int\!\left(
\frac12\Phi*\calQ\Phi+\frac13\,\Phi^3\right) .
\label{eq:calS}
\end{equation}
Our analysis proceeds as follows:
\begin{enumerate}
\item
We first present a set of six BV states $\{u_i(k)\}$
containing the tachyon state. This set of BV states is similar to
\eqref{eq:sixstates_long} for the photon longitudinal mode.

\item
We evaluate the matrix $\omega_{ij}(k)$ \eqref{eq:omega} for the six
BV states, and obtain the kinetic term of the action \eqref{eq:calS},
\begin{equation}
\calS_0[\Phi]=\frac12\int\!\Phi*\calQ\Phi ,
\label{eq:calS_0}
\end{equation}
by expanding $\Phi$ in terms of the six states.

\item
If the matrix $\omega_{ij}$ is non-degenerate, $\det\omega_{ij}\ne 0$,
we conclude that the tachyon field is an {\em unphysical} one.
On the other hand, if $\omega_{ij}$ is degenerate,
$\det\omega_{ij}=0$, and, furthermore, the kinetic term
\eqref{eq:calS_0} is reduced to \eqref{eq:L_chi}, the tachyon field is
a {\em physical} one.
\end{enumerate}
As a concrete choice of the six tachyon BV states $u_i(k)$, we take
\begin{alignat}{3}
\Ngh&=0:&\ u_0&=L \uu_0 R^{-1} ,
\nn\\
\Ngh&=1:&\ u_{1A}&=L \uu_{1A} R^{-1}+\xi\CR{\calPe}{L\uu_0 R^{-1}},
&\quad
u_{1B}&=L \uu_{1B} R^{-1}
-\left(1-\xi\right)\CR{\calPe}{L\uu_0 R^{-1}},
\nn\\
\Ngh&=2:&\ u_{2A}&=L \uu_{2A} R^{-1},
&\quad
u_{2B}&=\ACR{\calPe}{\left(1-\xi\right)L \uu_{1A} R^{-1}
+\xi L \uu_{1B} R^{-1}} ,
\nn\\
\Ngh&=3:&\ u_3&=i\CR{\calPe}{u_{2A}}
=i\CR{\calPe}{L \uu_{2A} R^{-1}} .
\label{eq:sixstates_MB}
\end{alignat}
Each state in \eqref{eq:sixstates_MB} consists of various ingredients.
First, $\uu_i(k)$ ($i=0,\,1A,\,1B,\,2A$) are defined by
\begin{align}
\uu_0&=-i\frac{B}{\Ke}\,\eaKe c V_k\,\eaKe ,
\nn\\
\uu_{1A}&=\eaKe c V_k\,\eaKe ,
\nn\\
\uu_{1B}&=\left(1-k^2\right)\frac{B}{\Ke}\,\eaKe cKc\,V_k\,\eaKe
+\frac{\veps}{\Ke}\,\eaKe c V_k\,\eaKe ,
\nn\\
\uu_{2A}&=\eaKe cKc V_k\,\eaKe .
\label{eq:uu}
\end{align}
By $\QB$, they are related by
\begin{equation}
i\QB\uu_0=\uu_{1A}-\uu_{1B},
\qquad
\QB\uu_{1A}=\QB\uu_{1B}=\left(1-k^2\right)\uu_{2A} .
\end{equation}
Note that there appear in \eqref{eq:uu} $\eaKe$ instead of $\eaK$.
Namely, each state in \eqref{eq:uu} is multiplied by an extra
factor $e^{-2\aw\veps}$. Though this is merely a c-number factor which
is reduced to one in the limit $\veps\to 0$, it makes the expressions
of various $O(1/\veps)$ quantities simpler as we will see
in Sec.\ \ref{sec:BV_2b}.

Second, $L=L(\Ke)$ and $R=R(\Ke)$ in \eqref{eq:sixstates_MB} are
functions of only $\Ke$.
Though they are quite arbitrary at this stage,
we will determine later, for each classical solution $\calPe$, their
small $\Ke$ behavior from the requirement that the EOM against
$u_{1A/B}$ and that against the commutators $\CR{u_0}{u_{1A/B}}$
hold.
Finally, $\xi$ in \eqref{eq:sixstates_MB} is a parameter
related to the arbitrariness in the definitions of $u_{1A}$ and
$u_{1B}$.

The action of the BRST operator $\calQ$ \eqref{eq:calQ} on the six
states of \eqref{eq:sixstates_MB} is given by
\begin{align}
i\calQ\,u_0&=u_{1A}-u_{1B} ,
\nn\\[2mm]
\calQ\Pmatrix{u_{1A}\\ u_{1B}}
&=\Pmatrix{1 \\ 1}\left[\left(1-k^2\right)u_{2A}+u_{2B}\right]
+i\Pmatrix{\xi \\ -1+\xi}\CR{\EOMe}{u_0} ,
\nn\\[2mm]
i\calQ\Pmatrix{u_{2A}\\ u_{2B}}
&=\Pmatrix{1\\ k^2-1}u_3+i\Pmatrix{0\\ 1}
\CR{\EOMe}{\left(1-\xi\right)u_{1A}+\xi u_{1B}} ,
\nn\\[2mm]
\calQ\,u_3&=i\CR{\EOMe}{u_{2A}} ,
\label{eq:calQu_i}
\end{align}
where $\EOMe$ is defined by
\begin{equation}
\EOMe=\QB\calPe+\calPe^2 ,
\label{eq:EOMe}
\end{equation}
and given explicitly by \eqref{eq:EOM_calPe}.

The set of six BV states \eqref{eq:sixstates_MB} has been constructed
by comparing its BRST transformation property \eqref{eq:calQu_i} with
that of the six BV states \eqref{eq:sixstates_long} for
the longitudinal photon and by taking into account that $\EOMe$ and
$\calPe$ are both apparently of $O(\veps)$.
First, $\uu_{1A}$ is the tachyon state on the unstable vacuum,
and $\uu_0$ is $\uu_{1A}$ multiplied by the ``homotopy operator''
$B/\Ke$ of $\QB$. We start with $u_0$, which is $\uu_0$ dressed by
$L$ and $R^{-1}$, and divided $\calQ u_0$ into the difference of
$u_{1A}$ and $u_{1B}$ as given by the first equation of
\eqref{eq:calQu_i}.
If we ignore the apparently of $O(\veps)$ terms, $u_{1A}$ is
the dressed tachyon state, and $u_{1B}$, which is multiplied by
$(1-k^2)$ vanishing at the tachyon on-shell $k^2=1$,
corresponds to $k^2 u_{1B}$ in the first equation of
\eqref{eq:QBu_long} for the massless longitudinal photon.
We have distributed $\CR{\calPe}{L\uu_0 R^{-1}}$ in $i\calQ u_0$ to
$u_{1A}$ and $u_{1B}$ with coefficients specified by the parameter
$\xi$.
Then, we consider $\calQ u_{1A}$ and $\calQ u_{1B}$, which are
equal to each other if $\calQ^2=0$ and hence $\EOMe=0$ holds
(see the second equation of \eqref{eq:calQu_i}).
We have chosen $u_{2A}$ and $u_{2B}$ as the part of $\calQ u_{1A}$
which is multiplied by $(1-k^2)$ and the rest, respectively.
For $u_{2A}$ and $u_{2B}$, we have no clear correspondence with the
BV states of the longitudinal photon.
Finally,  $u_3$ is naturally defined from
$\calQ\!\left(u_{2A},u_{2B}\right)$ as given in the last equation of
\eqref{eq:calQu_i}.

Our choice \eqref{eq:sixstates_MB} of the six states $u_i$ is of
course not a unique one. For instance, in \eqref{eq:uu}, the part
$\left(\veps/\Ke\right)\eaKe cV_k\eaKe$ in $\uu_{1B}$ may be moved to
$\uu_{1A}$ to replace $\uu_{1A}$, $\uu_{1B}$ and $\uu_{2A}$ in
\eqref{eq:sixstates_MB} with the following ones:
\begin{align}
\uu_{1A}&=\frac{K}{\Ke}\,\eaKe cV_k \,\eaKe ,
\nn\\
\uu_{1B}&=\left(1-k^2\right)\frac{B}{\Ke}\,\eaKe cKcV_k\,\eaKe ,
\nn\\
\uu_{2A}&=\frac{K}{\Ke}\,\eaKe cKc V_k\,\eaKe .
\label{eq:another_uu}
\end{align}
For $u_{2A}$ and $u_{2B}$, we may take more generic linear combinations
of the three terms $L \uu_{2A} R^{-1}$,
$\ACR{\calPe}{L \uu_{1A} R^{-1}}$ and
$\ACR{\calPe}{L \uu_{1B} R^{-1}}$.
However, here in this paper, we carry out the BV analysis by
adopting the states of \eqref{eq:sixstates_MB} with $\uu_i$ given by
\eqref{eq:uu}.
In this sense, our analysis is rather an ``experiment'' and is not a
comprehensive one.
We do not know whether the conclusion of tachyon being physical or
unphysical can be changed by taking another set of tachyon BV
states.\footnote{
In Sec.\ \ref{sec:summary}, we argue the stability of the
(un)physicalness of tachyon fluctuation under the change of the
parameter $\xi$ in \eqref{eq:sixstates_MB} and under the replacement
of $\uu_i$ \eqref{eq:uu} with those given by \eqref{eq:another_uu}.
}

\subsection{$\omega_{ij}^{(a,b)}(k)$}

As we will see later, $L(\Ke)$ and $R^{-1}(\Ke)$ appearing in the
definition of $u_i$ \eqref{eq:sixstates_MB} play a crucial role
in making the EOM terms to vanish on the 2-brane. However, the pair
$(L,R)$ is not uniquely determined by this requirement alone.
Therefore, we put a superscript $(a)$ on $(L,R)$ and the corresponding
states $u_i$ in \eqref{eq:sixstates_MB} to distinguish different
choices of $(L,R)$. For example, we write
\begin{equation}
u_0^{(a)}=\La\,\uu_0\left(1/\Ra\right) .
\label{eq:u_0^(a)}
\end{equation}
Then, the matrix $\omega_{ij}$ \eqref{eq:omega} now has another index
$(a,b)$:
\begin{equation}
\int\!u_i^{(a)}(k')\,u_j^{(b)}(k)=\omega_{ij}^{(a,b)}(k)
\times(2\pi)^{26}\delta^{26}(k'+k) ,
\label{eq:omega_ij^(a,b)}
\end{equation}
with $i,j=0,\,1A,\,1B,\,2A,\,2B,\,3$.
However, $\omega_{ij}^{(a,b)}(k)$ should not be regarded as a matrix
with its left index $(i,a)$ and right one $(j,b)$;
it is still a $6\times 6$ matrix with a fixed pair of $(a,b)$.
When we consider the action \eqref{eq:calS} in the final step of our
analysis, we put $(a)=(b)$ by taking a particular $(L,R)$.

We see that all the components of $\omega_{ij}^{(a,b)}$ are not
independent. By considering
$\int\!u_0^{(a)}i\calQ u_{2A}^{(b)}$,
$\int\!u_0^{(a)}i\calQ u_{2B}^{(b)}$ and
$\int\!u_{1B}^{(a)}\calQ u_{1A}^{(b)}$, and using
\begin{equation}
\int\!A_1*\calQ A_2=-(-1)^{A_1}\int\!(\calQ A_1)*A_2 ,
\label{eq:partint_calQ}
\end{equation}
and \eqref{eq:calQu_i}, we obtain the following relations:
\begin{align}
\omega_{1B,2A}^{(a,b)}&=\omega_{1A,2A}^{(a,b)}+\omega_{0,3}^{(a,b)} ,
\label{eq:rel_omegaI}
\\
\omega_{1B,2B}^{(a,b)}
&=\omega_{1A,2B}^{(a,b)}-\left(1-k^2\right)\omega_{0,3}^{(a,b)} ,
\label{eq:rel_omegaII}
\\
\left(1-k^2\right)\omega_{1B,2A}^{(a,b)}+\omega_{1B,2B}^{(a,b)}
&=\left(1-k^2\right)\omega_{1A,2A}^{(b,a)}+\omega_{1A,2B}^{(b,a)} .
\label{eq:rel_omegaIII}
\end{align}
In deriving the last two relations, we have assumed the vanishing
of the EOM terms:
\begin{equation}
\int\!\cR{u_0^{(a)}}{u_i^{(b)}}*\EOMe=0,
\qquad (i=1A,\,1B) .
\label{eq:CRu_0u_iEOMe=0}
\end{equation}
{}From \eqref{eq:rel_omegaI} and \eqref{eq:rel_omegaII}, we also
have
\begin{equation}
\left(1-k^2\right)\omega_{1B,2A}^{(a,b)}+\omega_{1B,2B}^{(a,b)}
=\left(1-k^2\right)\omega_{1A,2A}^{(a,b)}+\omega_{1A,2B}^{(a,b)} .
\label{eq:rel_omegaIV}
\end{equation}
Therefore, among the five components, $\omega_{1A/B,2A/B}^{(a,b)}$ and
$\omega_{0,3}^{(a,b)}$, we can choose $\omega_{1A,2A}^{(a,b)}$,
$\omega_{1A,2B}^{(a,b)}$ and $\omega_{0,3}^{(a,b)}$ as independent
ones, and write the submatrix $\Omega^{(a,b)}$ as
\begin{equation}
\Omega^{(a,b)}\equiv
\Pmatrix{
\omega_{1A,2A}^{(a,b)} &\omega_{1A,2B}^{(a,b)}
\\[3mm]
\omega_{1B,2A}^{(a,b)} &\omega_{1B,2B}^{(a,b)}
}
=\Pmatrix{
\omega_{1A,2A}^{(a,b)} &\omega_{1A,2B}^{(a,b)}
\\
\omega_{1A,2A}^{(a,b)}+\omega_{0,3}^{(a,b)}
&
\omega_{1A,2B}^{(a,b)}+\left(k^2-1\right)\omega_{0,3}^{(a,b)}
} .
\label{eq:Omega}
\end{equation}
Its determinant is given by
\begin{equation}
\abs{\Omega^{(a,b)}}
=\omega_{0,3}^{(a,b)}\left[\left(k^2-1\right)\omega_{1A,2A}^{(a,b)}
-\omega_{1A,2B}^{(a,b)}\right] .
\label{eq:absOmega}
\end{equation}
Using \eqref{eq:calQu_i} and assuming \eqref{eq:CRu_0u_iEOMe=0}, we
also obtain the following useful formulas:
\begin{align}
\int\!u_i^{(a)}\calQ u_j^{(b)}
&=\left(1-k^2\right)\omega_{1A,2A}^{(a,b)}+\omega_{1A,2B}^{(a,b)},
\qquad
(i,j=1A,\,1B) ,
\label{eq:int_u1calQu1}
\\
\int\!u_0^{(a)}i\calQ\,\bigl(u_{2A}^{(b)},u_{2B}^{(b)}\bigr)
&=\left(1,k^2-1\right)\omega_{0,3}^{(a,b)} ,
\label{eq:int_u0calQu2}
\end{align}
where we have omitted $(2\pi)^{26}\delta^{26}(k'+k)$ on the RHS.

\subsection{Formulas for the EOM tests and $\omega_{ij}^{(a,b)}$}
\label{sec:EOMtest+omega}

For the BV analysis for a given $\Ge$, we need to evaluate
(i) the EOM test of $\calPe$ against $u_{1A/B}$ and
$\cR{u_0^{(a)}}{u_{1A/B}^{(b)}}$,
and (ii)  $\omega_{0,3}^{(a,b)}$, $\omega_{1A,2A}^{(a,b)}$ and
$\omega_{1A,2B}^{(a,b)}$.
For $u_{1A}$ and $u_{1B}$ containing the parameter $\xi$ (see
\eqref{eq:sixstates_MB}), it is convenient to introduce
$w_{\ell}^{(a)}$ ($\ell=A,B,C$) defined by
\begin{equation}
\wA^{(a)}=\La\uu_{1A}\Ra{}^{-1},
\qquad
\wB^{(a)}=\La\uu_{1B}\Ra{}^{-1},
\qquad
\wC^{(a)}=\CR{\calPe}{\La\uu_0 \Ra{}^{-1}} ,
\label{eq:w_1ell}
\end{equation}
and express $u_{1A/B}^{(a)}$ as
\begin{equation}
u_{1A}^{(a)}=\wA^{(a)}+\xi_a\,\wC^{(a)},
\qquad
u_{1B}^{(a)}=\wB^{(a)}-\left(1-\xi_a\right)\wC^{(a)} ,
\end{equation}
where we have allowed the parameter $\xi$ to depend on the index
$a$ of $\La$ and $\Ra$.
For $\La(\Ke)$ and $\Ra(\Ke)$, we assume that their leading behaviors
for $\Ke\sim 0$ are
\begin{equation}
\La(\Ke)\sim \Ke^{m_a},\qquad \Ra(\Ke)\sim \Ke^{n_a},
\label{eq:LsimKe^m_a}
\end{equation}
and give their remaining $\Ke$-dependences as Laplace transforms:
\begin{equation}
\La(\Ke)=\Ke^{m_a}\int_0^\infty\!ds_a\,v_L^{(a)}(s_a)\,e^{-\Ke s_a},
\qquad
\frac{1}{\Ra(\Ke)}=\frac{1}{\Ke^{n_a}}\int_0^\infty\!d\sia
\,v_{1/R}^{(a)}(\sia)\,e^{-\Ke\sia} .
\label{eq:LandR_by_v}
\end{equation}
As given in \eqref{eq:LandR_by_v}, we adopt $s_a$ and $\sia$ as the
integration variable of the Laplace transform of $\La$ and $1/\Ra$,
respectively.
We adopt the following normalization for $v_L^{(a)}$ and
$v_{1/R}^{(a)}$:
\begin{equation}
\int_0^\infty\!ds_a\,v_L^{(a)}(s_a)
=\int_0^\infty\!d\sia\,v_{1/R}^{(a)}(\sia)=1 .
\label{eq:normalization}
\end{equation}
Namely, the coefficients of the leading terms \eqref{eq:LsimKe^m_a}
are taken to be equal to one.
The pair $(m_a,n_a)$ and the associated $v_L^{(a)}(s_a)$ and
$v_{1/R}^{(a)}(\sia)$ should be determined by the requirement of the
EOM as stated before.
Concerning the choice of $(m_a,n_a)$, it would be natural to consider
the  case $m_a=n_a$ since the overall order of the BV states $u_i$
\eqref{eq:sixstates_MB} with respect to $\Ke$ for $\Ke\sim 0$ is not
changed from the case without $\La$ and $\Ra$.
We will restrict ourselves to the case $m_a=n_a$ in the concrete
calculations given in Secs.\ \ref{sec:BV_2b} and \ref{sec:BV_TV}.

Then, the three kinds of quantities necessary for the BV analysis are
expressed as the following integrations over the Laplace transform
variables:
\begin{align}
\int\!w_{\ell}*\EOMe&=\int_0^\infty\!ds\,v_L(s)
\int_0^\infty\!d\si\,v_{1/R}(\si)\,E_{\ell}(s,\si) ,
\label{eq:def_E_1ell}
\\[2mm]
i\int\!\CR{w_{\ell}^{(a)}(k')}{u_0^{(b)}(k)}*\EOMe
&=\intfour\!\! E_{\ell,0}^{(a,b)}(s_a,s_b,\sia,\sib)
\times(2\pi)^{26}\delta^{26}(k'+k) ,
\label{eq:def_E_1ell-0}
\\[2mm]
\omega_{ij}^{(a,b)}(k)
&=\intfour\!\! W_{ij}^{(a,b)}(s_a,s_b,\sia,\sib) ,
\label{eq:def_W_ij}
\end{align}
where $\int_{(s_a,s_b,\sia,\sib)}$ is the integration defined by
\begin{equation}
\intfour=\int_0^\infty\!ds_a\,v_L^{(a)}(s_a)
\int_0^\infty\!ds_b\,v_L^{(b)}(s_b)
\int_0^\infty\!d\sia\,v_{1/R}^{(a)}(\sia)
\int_0^\infty\!d\sib\,v_{1/R}^{(b)}(\sib) .
\label{eq:intfour}
\end{equation}
The explicit expressions of $E_{\ell}$, $E_{\ell,0}^{(a,b)}$
($\ell=A,B,C$) and $W_{ij}^{(a,b)}$ are lengthy and hence are
summarized in Appendix \ref{app:EEW}.
They are given as sliver frame integrations containing a single or no
$B$. Though some of their defining expressions contain two or more
$B$, we have used the $\KBc$ algebra to reduce them to sliver frame
integrations with a single $B$.

The three $E_{\ell}$ are not independent, but they satisfy the
following relation:
\begin{equation}
\EA-\EB+\EC=0 .
\label{eq:E11-E12+E13=0}
\end{equation}
This follows from $\wA-\wB+\wC=u_{1A}-u_{1B}=i\calQ u_0$
(see the first of \eqref{eq:calQu_i}) and the Bianchi identity:
\begin{equation}
\calQ\,\EOMe=0 .
\end{equation}
Eq.\ \eqref{eq:E11-E12+E13=0} can be used as a consistency check of
the calculations.

\subsection{The action of the fluctuation in the non-degenerate case}
\label{sec:calS0_nondegen}

Let us consider the kinetic term $\calS_0[\Phi]$ \eqref{eq:calS_0}
in the case of non-degenerate $\omega_{ij}$.
We have attached the superscript $(a,b)$ on $\omega_{ij}^{(a,b)}$ for
distinguishing $\left(v_L,v_{1/R},\xi\right)$ defining
the state $u_i$ and that defining $u_j$.
However, when we express the fluctuation in terms of the basis
$\{u_i\}$ and the corresponding component fields, we choose one
particular $\left(v_L,v_{1/R},\xi\right)$.
Namely, when we consider the action \eqref{eq:calS_0}, there appear
only $\omega_{ij}^{(a,a)}$ with $(a)=(b)$.
Therefore, we here omit the superscript $(a,a)$ and simply write
$\omega_{ij}$.

When $\omega_{ij}$ is non-degenerate and the determinant
\eqref{eq:absOmega} is not identically equal to zero,
$\abs{\Omega}\ne 0$, it is convenient to move to the Darboux basis
by switching from $(u_{2A},u_{2B})$ to $(u_{2P},u_{2Q})$ defined by
\begin{equation}
\left(u_{2P},u_{2Q}\right)=\left(u_{2A},u_{2B}\right)\Omega^{-1} ,
\label{eq:u2P_u2Q}
\end{equation}
where the inverse matrix $\Omega^{-1}$ is given by
\begin{equation}
\Omega^{-1}=\frac{1}{\abs{\Omega}}\Pmatrix{
\omega_{1B,2B} & -\omega_{1A,2B}\\
-\omega_{1B,2A} & \omega_{1A,2A}},
\qquad
\abs{\Omega}=\omega_{0,3}\left[\left(k^2-1\right)\omega_{1A,2A}
-\omega_{1A,2B}\right] .
\label{eq:Omega^-1}
\end{equation}
The new set $\{u_0,u_{1A},u_{1B},u_{2P},u_{2Q},u_3\}$ is in fact
a Darboux basis since we have
\begin{equation}
\Pmatrix{\omega_{1A,2P} & \omega_{1A,2Q}\\
\omega_{1B,2P} & \omega_{1B,2Q}}
=\int\Pmatrix{u_{1A}\\ u_{1B}}\left(u_{2P}, u_{2Q}\right)
=\Pmatrix{1 & 0 \\ 0 & 1} .
\end{equation}
Instead of \eqref{eq:int_u0calQu2}, $(u_{2P},u_{2Q})$ satisfies
\begin{equation}
\int\!u_0\,i\calQ\left(u_{2P},u_{2Q}\right)
=\omega_{0,3}\left(1,k^2-1\right)\Omega^{-1}
=(-1,1) .
\end{equation}

For expressing $\Phi$ in terms of the Darboux basis, it is more
convenient to use still another one $\{\wt{u}_i\}$ with tilde, which
is defined by multiplying the states corresponding to the fields and
anti-fields by $\sqrt{\varpi(k)}$ and its inverse, respectively:
\begin{equation}
\Pmatrix{\wt{u}_0 \\ \wt{u}_{1A} \\ \wt{u}_{2Q}}
=\sqrt{\varpi(k)}\Pmatrix{u_0 \\ u_{1A} \\ u_{2Q}},
\qquad
\Pmatrix{\wt{u}_{1B} \\ \wt{u}_{2P} \\ \wt{u}_3}
=\frac{1}{\sqrt{\varpi(k)}}\Pmatrix{u_{1B} \\ u_{2P} \\ u_3} ,
\label{eq:wtu}
\end{equation}
with $\varpi(k)$ given by
\begin{equation}
\varpi(k)=\left(k^2-1\right)\omega_{1A,2A}(k)-\omega_{1A,2B}(k) .
\label{eq:varpi}
\end{equation}
Then, expressing $\Phi$ as
\begin{equation}
\Phi=\!\int_k\!\Bigl\{
\wt{u}_0(k)\,C(k)+\wt{u}_{1A}(k)\,\chi(k)+\wt{u}_{1B}(k)\,\ol{C}_\star(k)
+\wt{u}_{2P}(k)\,\chi_\star(k)+\wt{u}_{2Q}(k)\,\ol{C}(k)
+\wt{u}_3(k)\,C_\star(k)
\Bigr\} ,
\end{equation}
the kinetic term \eqref{eq:calS_0} is given in terms of the component
fields and anti-fields as
\begin{align}
\calS_0[\Phi]
&=\int_k\biggl\{-\frac12\left(\varpi\,\chi+\ol{C}_\star\right)\!(-k)
\left(\varpi\,\chi+\ol{C}_\star\right)\!(k)
+i\left(\varpi\,\ol{C}-\chi_\star\right)\!(-k)\,C(k)\biggr\} .
\label{eq:calS0_nondegen}
\end{align}
The action \eqref{eq:S0_long} for the photon longitudinal mode
on the unstable vacuum is essentially the special case of
\eqref{eq:calS0_nondegen} with $\varpi(k)=k^2$, and the gauge-fixing
process for \eqref{eq:calS0_nondegen} goes in the same manner as for
\eqref{eq:S0_long}. Adopting the gauge $L$ with
$\chi_\star=\ol{C}_\star=C_\star=0$, the gauge-fixed action and the
BRST transformation are given by
\begin{equation}
\wh{\calS}_0=\calS_0\bigr|_L
=\int_k\biggl\{-\frac12\left(\varpi\,\chi\right)\!(-k)
\left(\varpi\,\chi\right)\!(k)
+i\left(\varpi\,\ol{C}\right)\!(-k)\,C(k)\biggr\} ,
\label{eq:whcalS0}
\end{equation}
and
\begin{align}
\whdB\chi(k)&=i\left.\fDrv{\calS_0}{\chi_\star(-k)}\right|_L=C(k) ,
\nn\\
\whdB\ol{C}(k)&=i\left.\fDrv{\calS_0}{\ol{C}_\star(-k)}\right|_L
=-i\varpi(k)\chi(k) ,
\nn\\
\whdB C(k)
&=i\omega_{0,3}(k)^{-1}\left.\fDrv{\calS_0}{C_\star(-k)}\right|_L
=0 .
\end{align}
If $\varpi(k)$ has a zero at $k^2=-m^2$, the action \eqref{eq:whcalS0}
describes a totally unphysical system with mass $m$ explained in the
Introduction.

The above argument leading to \eqref{eq:calS0_nondegen} does not apply
if we $\omega_{ij}$ is degenerate.
In such a case, the system can describe a physical one in general.

\subsection{(Non-)hermiticity of the BV states}
\label{sec:hermiticity}

Our tachyon BV basis $\{u_i\}$ given by \eqref{eq:sixstates_MB} has in
fact a problem that it does not satisfy the hermiticity condition.
We will explain it in this subsection.

In the original CSFT action \eqref{eq:S}, the the string field $\Psi$
is assumed to be hermitian; $\Psi^\dagger=\Psi$, or more generally,
$\Psi^\dagger=W(K)\Psi W(K)^{-1}$ with $W(K)$ depending only on $K$.
This constraint ensures the reality of the action \eqref{eq:S} and, at
the same time, prevents the duplication of each fluctuation modes.
Then, let us consider the hermiticity for the action
\eqref{eq:calS} of the fluctuation $\Phi$ around $\calPe$.
First, $\calPe$ \eqref{eq:calPe} satisfies the hermiticity
in the following sense:
\begin{equation}
\calPe^\dagger
=\veps\times\left(1-\Ge\right)c\,\Ge Bc\frac{1}{\Ge}
=W\calPe W^{-1} ,
\end{equation}
with $W$ given by
\begin{equation}
W=\Ge\left(1-\Ge\right) .
\label{eq:W_calPe}
\end{equation}
Therefore, the fluctuation $\Phi$ in \eqref{eq:calS} must satisfy the
same hermiticity:
\begin{equation}
\Phi^\dagger=W\Phi W^{-1} .
\label{eq:Herm_Phi}
\end{equation}
If \eqref{eq:Herm_Phi} holds, it follows that $\calQ\Phi$ with $\calQ$
defined by \eqref{eq:calQ} also satisfies the same
hermiticity\footnote{
In deriving \eqref{eq:Herm_calQPhi}, we use the property
$\QB K=0$ and
$\bigl(\QB A\bigr)^\dagger=-(-1)^A\QB A^\dagger$
valid for any string field $A$.
}
\begin{equation}
\bigl(\calQ\Phi\bigr)^\dagger=W\left(\calQ\Phi\right)W^{-1} ,
\label{eq:Herm_calQPhi}
\end{equation}
and hence that the action \eqref{eq:calS} is real.
In the expansion \eqref{eq:Phi=sumu_ivarphi^i} of $\Phi$
in terms of the basis $\{u_i(k)\}$ and the component fields
$\varphi^i(k)$,
the hermiticity of $\Phi$, \eqref{eq:Herm_Phi}, is realized by imposing
\begin{equation}
u_i(k)^\dagger=W\,u_i(-k)\, W^{-1} ,
\label{eq:Herm_u_i}
\end{equation}
and $\varphi^i(k){}^\dagger=\varphi^i(-k)$.
However, our BV states \eqref{eq:sixstates_MB} do not satisfy this
hermiticity condition.

One way to realize the hermiticity \eqref{eq:Herm_u_i} is to take,
instead of the states $u_i$ \eqref{eq:sixstates_MB}, the following
ones $U_i$:
\begin{equation}
U_i(k)=\frac12\left[u_i(k)+W^{-1}\,u_i(-k)^\dagger W\right] .
\label{eq:U_i}
\end{equation}
In fact, $U_i(k)$ satisfies \eqref{eq:Herm_u_i} since $W$
\eqref{eq:W_calPe} is hermitian, $W^\dagger=W$.
The relations \eqref{eq:calQu_i} under the operation of $\calQ$ remain
valid when $u_i$ is replaced with $U_i$.

However, the results of the BV analysis which will be presented in
Secs.\ \ref{sec:BV_2b} and \ref{sec:BV_TV} are largely changed if we
adopt the hermitian basis $\{U_i\}$ instead of $\{u_i\}$.
The EOM against $U_{1A/B}$ is the same as that for $u_{1A/B}$.
On the other hand, the cross terms among the two terms on the RHS of
\eqref{eq:U_i} are added to the EOM against the commutator
$\cR{U_{1A/B}^{(a)}}{U_0^{(b)}}$
as well as $\omega_{ij}^{(a,b)}$ defined by \eqref{eq:omega_ij^(a,b)}
with $u_i$ replaced with $U_i$.
Sample calculations show that these cross terms change the results of
Secs.\ \ref{sec:BV_2b} and \ref{sec:BV_TV} to much more complicated
ones.
For example, the EOM against $\cR{U_{1A/B}^{(a)}}{U_0^{(b)}}$
on the 2-brane no longer holds for any
$\left(v_L,v_{1/R},\xi\right)$ with $(m,n)=(1,1)$.
Therefore, we will continue our analysis by using the original
non-hermitian basis $\{u_i\}$, though this is certainly a problem to
be solved in the future.

\section{BV analysis around the 2-brane solution}
\label{sec:BV_2b}

In this section, we carry out the BV analysis of the six states of
\eqref{eq:sixstates_MB} for the 2-brane solution given by
$G_\textrm{2b}$ in \eqref{eq:twoG}.
Our analysis consists of the following three steps:
\begin{enumerate}
\item
Evaluation of the EOM of $\calPe$ against $u_{1A/B}$ and
$\cR{u_0^{(a)}}{u_{1A/B}^{(b)}}$ (recall \eqref{eq:CRu_0u_iEOMe=0} for
the necessity of the latter).
{}From the vanishing of these EOMs, we determine the allowed set of
$(L(\Ke),R(\Ke))$.

\item
Calculation of $\omega_{0,3}$, $\omega_{1A,2A}$ and $\omega_{1A,2B}$
for $(L(\Ke),R(\Ke))$ determined above.
Our expectation is that $\omega_{0,3}=0$, namely, that the present set
of six BV states is degenerate and therefore the tachyon can be
physical.

\item
Derivation of the kinetic term $\calS_0[\Phi]$ \eqref{eq:calS_0} of
the fluctuation $\Phi$ in terms of the component fields defined by the
basis $\{u_i\}$.
\end{enumerate}

\subsection{EOM against  $u_{1A}$ and $u_{1B}$}
First, let us consider the EOM test against $u_{1A}$ and $u_{1B}$.
For this purpose, we have to evaluate $E_{\ell}(s,\si)$ of
\eqref{eq:def_E_1ell}, which is given by
\eqref{eq:E_11}--\eqref{eq:E_13} for a generic $\Ge$.
For the 2-brane solution with $G=G_\textrm{2b}$ \eqref{eq:twoG},
$E_{\ell}$ are given explicitly by
\begin{align}
\EA&=-\veps\int\!Bc\Ke^{m-1}
e^{-(\aw+s)\Ke}c\,\frac{\Ke^{1-n}}{1+\Ke}\,e^{-(\aw+\si)\Ke}
c\cCR{\Ke}{1+\invKe} ,
\label{eq:E_11_2b}
\\
\EB&=-\veps\int\!BcKc\,\frac{\Ke^{1-n}}{1+\Ke}\,e^{-(\aw+\si)\Ke}
\,c\cCR{\Ke}{1+\invKe}\frac{e^{-(\aw+s)\Ke}}{\Ke^{2-m}}
\nn\\
&\qquad
-\veps^2\int\!Bc\frac{e^{-(\aw+s)\Ke}}{\Ke^{2-m}}
\,c\,\frac{\Ke^{1-n}}{1+\Ke}\,e^{-(\aw+\si)\Ke}c
\cCR{\Ke}{1+\invKe},
\label{eq:E_12_2b}
\\
\EC&=\veps^2\int\!Bc\,\frac{\Ke^{1-n}}{1+\Ke}\,e^{-(\aw+\si)\Ke}
c\cCR{K}{\invKe}c\,\frac{e^{-(\aw+s)\Ke}}{\Ke^{2-m}} \ .
\label{eq:E_13_2b}
\end{align}
where $\cCR{\Ke}{1+(1/\Ke)}$ is defined by (see \eqref{eq:cCRKeGe})
\begin{equation}
\cCR{\Ke}{1+\invKe}=\Ke\,c\left(1+\invKe\right)
-\left(1+\invKe\right)c\,\Ke .
\end{equation}
Since the order of the correlator $\int\!Bc\Ke^p c\Ke^q c\Ke^r c\Ke^s$
with respect to $\veps$ is
$O\!\left(\veps^{\min(p+q+r+s-3,0)}\right)$,\footnote{
This formula is derived by using the scaling property
$G(\lambda t_1,\lambda t_2,\lambda t_3,\lambda t_4)
=\lambda^3 G(t_1,t_2,t_3,t_4)$ of the correlator
$G(t_1,t_2,t_3,t_4)=
\int\!Bc\,e^{-t_1 K}c\,e^{-t_2 K}c\,e^{-t_3 K}c\,e^{-t_4 K}$
given by \eqref{eq:Bcccc_col} and \eqref{eq:VEVBcccc}.
For $p+q+r+s\ge 4$, the correlator contains divergences from
$K=\infty$ and hence not regularized by $\veps$. This is the reason
why ``min'' appears in the formula.
}
we find that
\begin{align}
\EA&=O\!\left(\veps^{1+\min(m-n-3+\delta_{m,1}+\delta_{n,1}
+\delta_{m,1}\delta_{n,1},0)}\right) ,
\nn\\
\EB&=O\!\left(\veps^{1+\min(m-n-3+2\delta_{n,1}+\delta_{n,0},0)}\right)
+O\!\left(\veps^{2+\min(m-n-4+\delta_{m,2}+\delta_{n,1}
+\delta_{m,2}\,\delta_{n,1},0)}\right) ,
\nn\\
\EC&=O\!\left(\veps^{2+\min(m-n-4+\delta_{n,1},0)}\right) ,
\label{eq:E_ell_2b_order}
\end{align}
where the Kronecker-delta terms are due to the identities
$c^2=cKcKc=0$, and the two terms in $\EB$ corresponds to those in
\eqref{eq:E_12_2b}.
For a given $(m,n)$, $E_{\ell}$ can be evaluated by using the
formulas of the $Bcccc$ correlators given in Appendix \ref{app:KBc}.
However, we cannot carry out the calculation for a generic $(m,n)$,
and the calculation for each $(m,n)$ is very cumbersome.
Therefore, we have evaluated $E_{\ell}$ only for two cases,
$(m,n)=(1,1)$ and $(0,0)$.
We have chosen $(m,n)=(1,1)$ since, as seen from
\eqref{eq:E_ell_2b_order}, $E_\ell$ are least singular with respect to
$\veps$ for $(m,n)=(1,1)$ if we restrict ourselves to the case
$m=n$.\footnote{
As we mentioned in Sec.\ \ref{sec:EOMtest+omega}, the case $m=n$ is
natural in the sense that the overall order of each BV state $u_i$
\eqref{eq:sixstates_MB} with respect to $\Ke$ is not changed by
$(L,R)$.
}
We have taken the other one $(m,n)=(0,0)$ including the simplest
case $L=R=1$ as a reference.

\subsubsection{$(m,n)=(1,1)$}

In this case, $E_{\ell}$ are given up to $O(\veps)$ terms by
\begin{align}
\EA(s,\si)&=0 ,
\nn\\
\EB(s,\si)
&=\EC(s,\si)
=-\frac{3}{\pi^2}C_1(\si)\times\invveps+\frac12\,C_2(\si) ,
\label{eq:E_1ell_2b_1-1}
\end{align}
where $C_{1,2}(\si)$ are defined by
\begin{equation}
C_1(\si)=\si+\aw+1,
\qquad
C_2(\si)=\left(\si+\aw+1\right)^2+1 .
\label{eq:C_1,2}
\end{equation}
We defer further arguments on the EOM against $u_{1A/B}$ for
$(m,n)=(1,1)$ till we discuss the EOM against
$\cR{u_{1A/B}^{(a)}}{u_0^{(b)}}$ in Sec.\
\ref{sec:E_1ell-0_2b_11-11}.

\subsubsection{$(m,n)=(0,0)$}
\label{sec:E_1ell_2b_00}

In this case, $E_{\ell}$ are all of $O(1/\veps^2)$:
\begin{equation}
\left(\EA,\EB,\EC\right)
=(1,2,1)\times\frac{3}{\pi^2}\frac{1}{\veps^2}
+O\!\left(\invveps\right) .
\label{eq:E_1ell_2b_0-0_e^-2}
\end{equation}
This implies that we have to choose $\xi=-1$ to make the
$1/\veps^2$ part of the EOM test against $u_{1A}$ and $u_{1B}$ to
vanish. Namely, we have to take
\begin{align}
u_{1A}=\wA-\wC ,
\qquad
u_{1B}=\wB-2\,\wC .
\label{eq:u_1A/B_0-0}
\end{align}
Then, the combinations of $E_{\ell}$ relevant to $u_{1A/B}$
are given as follows:
\begin{align}
\EA-\EC&=\EB-2\EC
=\left[\left(\frac{2}{\pi^2}+\frac53\right)C_1(\si)
+\frac{3}{\pi^2}\,D_1(s)\right]\frac{1}{\veps}
-\frac72\,C_2(\si)+\left(2-D_1(s)\right)C_1(\si) ,
\label{eq:E_1ell_2b_0-0}
\end{align}
with $D_1(s)$ defined by
\begin{equation}
D_1(s)=s+\aw .
\end{equation}

\subsection{EOM against  $\cR{u_{1A/B}^{(a)}}{u_0^{(b)}}$}

Next, we evaluate the EOM test against
$\cR{u_{1A/B}^{(a)}}{u_0^{(b)}}$,
namely, $E_{\ell,0}^{(a,b)}$ of \eqref{eq:def_E_1ell-0} given
explicitly by \eqref{eq:E_11_0}--\eqref{eq:E_13_0} for a generic
$\Ge$. As in the previous subsection, we consider only the two cases;
$(m_a,n_a)=(m_b,n_b)=(1,1)$ and $(0,0)$.
We will explain the calculations in the case of $(1,1)$ in great
detail. The same method will be used also in the calculation of
$\omega^{(a,b)}_{ij}$.

\subsubsection{$(m_a,n_a)=(m_b,n_b)=(1,1)$}
\label{sec:E_1ell-0_2b_11-11}

For $G=G_{2b}$ \eqref{eq:twoG} and for
$(m_a,n_a)=(m_b,n_b)=(1,1)$, $E_{A,0}^{(a,b)}$ \eqref{eq:E_11_0}
reads
\begin{align}
&E_{A,0}^{(a,b)}(s_a,s_b,\sia,\sib)
\nn\\
&=\veps\int\!Bc\,V_k
\frac{e^{-(\aw+\sib)\Ke}}{1+\Ke}
c\cCR{\Ke}{1+\invKe}\CR{c}{e^{-(\aw+s_a)\Ke}}
V_{-k}\frac{e^{-(2\aw+\sia+s_b)\Ke}}{\Ke}
\nn\\
&\quad
-\veps\int\!Bc\,V_k e^{-(2\aw+\sib+s_a)\Ke}
c\,V_{-k}\frac{e^{-(\aw+\sia)\Ke}}{1+\Ke}
\,c\!\cCR{\Ke}{1+\invKe}\!\!\frac{e^{-(\aw+s_b)\Ke}}{\Ke} .
\label{eq:E_11-0_2b}
\end{align}
Let us explain how we evaluate \eqref{eq:E_11-0_2b} and other
$E_{\ell,0}^{(a,b)}$ for a generic momentum $k_\mu$.
Let us consider, as an example, the contribution of the
$c\,e^{-(\aw+s_a)K}$ term of the commutator $\CR{c}{e^{-(\aw+s_a)K}}$
to the first integral of \eqref{eq:E_11-0_2b}:
\begin{align}
&\veps\int\!Bc\,V_k\frac{e^{-(\aw+\sib)\Ke}}{1+\Ke}
c\left(Kc\invKe-\invKe cK\right)c\,e^{-(\aw+s_a)\Ke}
V_{-k}\frac{e^{-(2\aw+\sia+s_b)\Ke}}{\Ke}
\nn\\
&=\eveps\,
\veps\!\int_0^\infty\!\!dt_1\,e^{-\left(1+\veps\right)t_1}\!
\int_0^\infty\!\!dt_2\,e^{-\veps t_2}\!
\int_0^\infty\!\!dt_3\,e^{-\veps t_3} F(t_1,t_2,t_3) ,
\label{eq:ex_E_11-0_2b}
\end{align}
with $\eveps$ defined by
\begin{equation}
\eveps=e^{-\veps\left(\pi+s_a+s_b+\sia+\sib\right)} .
\label{eq:eveps}
\end{equation}
In \eqref{eq:ex_E_11-0_2b}, $t_1$, $t_2$ and $t_3$ are the Schwinger
parameters for $1/(1+\Ke)$ and the two $1/\Ke$, respectively, and
the function $F(t_1,t_2,t_3)$ is given by
\begin{align}
&F(t_1,t_2,t_3)=-\left.\Drv{}{w_2}
G\bigl(t_1+\aw+\sib,w_2,t_2,t_3+3\aw+s_a+\sia+s_b;t_3+2\aw+\sia+s_b
\bigr)\right|_{w_2=0}
\nn\\
&\quad
+\left.\Drv{}{w_3}
G\bigl(t_1+\aw+\sib,t_2,w_3,t_3+3\aw+s_a+\sia+s_b;t_3+2\aw+\sia+s_b
\bigr)\right|_{w_3=0} ,
\end{align}
where $G$ is the product of the ghost correlator and the matter one on
the infinite cylinder of circumference $\ell=w_1+w_2+w_3+w_4$:
\begin{equation}
G(w_1,w_2,w_3,w_4;w_X)=\bigl\langle
Bc(0)c(w_1)c(w_1+w_2)c(w_1+w_2+w_3)\bigr\rangle_\ell
\times\abs{\frac{\ell}{\pi}\sin\frac{\pi w_X}{\ell}}^{-2k^2} .
\label{eq:G}
\end{equation}
The explicit expressions of the correlators are given in Appendix
\ref{app:KBc}.

One way to evaluate \eqref{eq:ex_E_11-0_2b} in the limit $\veps\to 0$
is to (i) make a change of integration variables from $(t_2,t_3)$ for
$1/\Ke$ to $(u,x)$ by $(t_2,t_3)=(u/\veps)\left(x,1-x\right)$,
(ii) carry out the $x$-integration first,
(iii) Laurent-expand the integrand in powers of $\veps$ to a
necessary order, and finally (iv) carry out the integrations over $u$
and $t_1$.
In fact, we obtained the results \eqref{eq:E_1ell_2b_1-1},
\eqref{eq:E_1ell_2b_0-0_e^-2} and \eqref{eq:E_1ell_2b_0-0} by this
method. However, it is hard to carry out explicitly the
$x$-integration in \eqref{eq:ex_E_11-0_2b} before Laurent-expanding
with respect to $\veps$ due to the presence of the $k^2$-dependent
matter correlator in \eqref{eq:G}.
On the other hand, Laurent-expanding the $(t_1,u,x)$-integrand with
respect to $\veps$ before carrying out the $x$-integration
sometimes leads to a wrong result.
Namely, the integration regions where $x$ or $1-x$ are of $O(\veps)$
can make non-trivial contributions.

Our manipulation for obtaining the correct result for
\eqref{eq:ex_E_11-0_2b} is as follows.
Eq.\ \eqref{eq:ex_E_11-0_2b}, which is multiplied by $\veps$, can be
non-vanishing due to negative powers of $\veps$ arising from the two
$1/\Ke$ at the zero eigenvalue $K=0$.
In the RHS of \eqref{eq:ex_E_11-0_2b}, this contribution
comes from any of the following three regions of the
$(t_2,t_3)$-integration:
\begin{alignat}{3}
\textrm{Region I:}&\quad &t_2&=\textrm{finite},\quad &t_3&\to\infty ,
\nn\\
\textrm{Region II:}&\quad &t_2&\to\infty,\quad &t_3&\to\infty ,
\nn\\
\textrm{Region III:}&\quad &t_2&\to\infty,
\quad &t_3&=\textrm{finite} .
\label{eq:ThreeRegions}
\end{alignat}
Concretely, \eqref{eq:ex_E_11-0_2b} is given as the sum of the
contributions from the three regions:
\begin{equation}
\eveps\int_0^\infty\!dt_1\,e^{-\left(1+\veps\right)t_1}
\left[(\textrm{I})+(\textrm{II})+(\textrm{III})\right] ,
\label{eq:t1int_I+II+III}
\end{equation}
with each term given by
\begin{align}
\textrm{(I)}&=
\int_\veps^\infty\!du\,e^{-u}\int_0^{\zeta u/\veps}\!dy\,
\Ser_\veps F\bigl(t_1,t_2=y,t_3=(u/\veps)-y\bigr) ,
\label{eq:(I)}
\\
\textrm{(II)}&=
\int_\veps^\infty\!du\,e^{-u}\int_\zeta^{1-\eta}\!dx\,\Ser_\veps
\frac{u}{\veps}F\bigl(t_1,t_2=xu/\veps,t_3=(1-x)u/\veps\bigr) ,
\label{eq:(II)}
\\
\textrm{(III)}&=
\int_\veps^\infty\!du\,e^{-u}\int_0^{\eta u/\veps}\!dy\,
\Ser_\veps F\bigl(t_1,t_2=(u/\veps)-y,t_3=y\bigr) ,
\label{eq:(III)}
\end{align}
where $\Ser_\veps$ denotes the operation of Laurent-expanding the
function with respect to $\veps$ to a necessary order.\footnote{
In this Laurent expansion, we treat $u$, $y$, $x$ and $1-x$ as
quantities of $O(\veps^0)$.
}
In each region, we have put $t_2+t_3=u/\veps$ and limited the
integration region of $u$ to $(\veps,\infty)$ since the other region
$(0,\veps)$ cannot develop a negative power of $\veps$.
As given in \eqref{eq:(I)}--\eqref{eq:(III)}, the three regions of
\eqref{eq:ThreeRegions} are specified by two parameters, $\zeta$ and
$\eta$, which we assume to be of $O(\veps^0)$.
Explicitly, the evaluation of the terms (I)--(III) goes as follows:\\
\noindent
\underline{Term (I)}

For \eqref{eq:(I)}, the Laurent expansion gives
\begin{equation}
\Ser_\veps F=\frac{2\pi^2}{3}\left(\frac{\veps}{u}\right)^3
\left(t_1+\sib+\aw\right)
\left(t_1+\sib+\aw+y\right)
\left(t_1+s_a+\sib+2\aw+y\right)^{-2k^2}y^3
+\ldots\ .
\label{eq:SerF_I}
\end{equation}
The leading term of the $y$-integration is of order
$(u/\veps)^{\max(5-2k^2,0)}$, and we obtain
\begin{equation}
(\textrm{I})\sim\int_\veps^\infty\!du\,e^{-u}
\left(\frac{\veps}{u}\right)^{3-\max(5-2k^2,0)}
=O\bigl(\veps^{\min(2k^2-2,1)}\bigr) ,
\label{eq:Eval_I}
\end{equation}
where we have used that\footnote{
Precisely, the RHS of \eqref{eq:intu_formula1} for $g=1$ should read
$O\!\left(\veps\ln\veps\right)$.

}
\begin{equation}
\int_\veps^\infty\!du\,e^{-u}\left(\frac{\veps}{u}\right)^g
=O\bigl(\veps^{\min(g,1)}\bigr) .
\label{eq:intu_formula1}
\end{equation}
The subleading term of \eqref{eq:SerF_I}, which is of
$O\!\left((\veps/u)^4\right)$, gives terms of order
$\veps^{\min(2k^2-2+p,1)}$ with $p=1,2,\cdots$.

\noindent
\underline{Term (II)}

The Laurent expansion in \eqref{eq:(II)} gives
\begin{equation}
\Ser_\veps \frac{u}{\veps}\,F=
2\left(t_1+\sib+\aw\right)
\frac{\pi x\cos\pi x-\sin\pi x}{\sin\pi x}
\left(\frac{u\sin\pi x}{\pi\veps}\right)^{2-2k^2}
+\ldots\ .
\label{eq:Ser(u/veps)F}
\end{equation}
Since the $x$-integration in the range $\zeta\le x\le 1-\eta$ is
finite, we obtain
\begin{equation}
(\textrm{II})\sim
\int_\veps^\infty\!du\,e^{-u}\left(\frac{u}{\veps}\right)^{2-2k^2}
=O\bigl(\veps^{\min(2k^2-2,1)}\bigr) .
\label{eq:Eval_II}
\end{equation}

\noindent
\underline{Term (III)}

The Laurent expansion in \eqref{eq:(III)} gives
\begin{equation}
\Ser_\veps F=-2\left(\sib+t_1+\aw\right)
\left(s_b+\sia+2\aw+y\right)^{-2k^2}
\left(s_a+s_b+\sia+3\aw+y\right)
+\ldots\ .
\end{equation}
Carrying out the $y$-integration, we get
\begin{align}
(\textrm{III})&=\frac{t_1+\sib+\aw}{(k^2-1)(2k^2-1)}
\int_\veps^\infty\!du\,e^{-u}\biggl[
\left(s_b+\sia+2\aw+y\right)^{1-2k^2}
\nn\\
&\qquad\times
\Bigl\{2(k^2-1)(s_a+\aw)+(2k^2-1)(s_b+\sia+2\aw+y)
\Bigr\}\biggr]^{y=\eta u/\veps}_{y=0}
\nn\\
&=-\frac{(t_1+\sib+\aw)\left(s_b+\sia+2\aw\right)^{1-2k^2}
}{(k^2-1)(2k^2-1)}
\Bigl\{2(k^2-1)(s_a+\aw)+(2k^2-1)(s_b+\sia+2\aw)\Bigr\}
\nn\\
&\qquad
+O\bigl(\veps^{\min(2k^2-2,1)}\bigr) ,
\label{eq:Eval_III}
\end{align}
where the last term is the contribution of the $y=\eta u/\veps$ term.

Summing the three terms, \eqref{eq:Eval_I}, \eqref{eq:Eval_II} and
\eqref{eq:Eval_III}, and carrying out the $t_1$-integration of
\eqref{eq:t1int_I+II+III}, we finally find that
\eqref{eq:ex_E_11-0_2b} is given by
\begin{equation}
-\frac{(\sib+\aw+1)\left(s_b+\sia+2\aw\right)^{1-2k^2}}{(k^2-1)(2k^2-1)}
\Bigl\{2(k^2-1)(s_a+\aw)+(2k^2-1)(s_b+\sia+2\aw)\Bigr\}
+O\bigl(\veps^{\min(2k^2-2,1)}\bigr) .
\end{equation}
This result can also be checked by numerically carrying out the
integrations of \eqref{eq:ex_E_11-0_2b} for given values of
$\veps$, $k^2$ and other parameters in \eqref{eq:ex_E_11-0_2b}.

The evaluation of the other term of the first integral of
\eqref{eq:E_11-0_2b}, namely, the term containing the
$e^{-(\aw+s_a)K}c$ part of the commutator $\CR{c}{e^{-(\aw+s_a)}}$, is
quite similar. In fact, the two terms of the commutator almost cancel
one another, and the whole of the first integral of
\eqref{eq:E_11-0_2b} turns out to be simply of
$O\bigl(\veps^{\min(2k^2-1,1)}\bigr)$.\footnote{
The actual $\veps$-dependence may be a milder one since we are not
taking into account the possibility of cancellations among the three
terms \eqref{eq:(I)}--\eqref{eq:(III)} for the whole of the first
integral of \eqref{eq:E_11-0_2b}.
In fact, numerical analysis supports a milder behavior
$O\bigl(\veps^{\min(2k^2,1)}\bigr)$.
}

Next, the second integral of \eqref{eq:E_11-0_2b} is given by
\begin{align}
&\eveps\,\veps\!\int_0^\infty\!\!dt_1\,e^{-(1+\veps)t_1}\!
\int_0^\infty\!\!dt_2\,e^{-\veps t_2}\Bigl\{
G(2\aw+\sib+s_a,t_1+\aw+\sia,t_2,\aw+s_b;2\aw+\sib+s_a)
\nn\\
&\quad
+(1+t_3)\left.\p_{w_3}G(2\aw+\sib+s_a,t_1+\aw+\sia,w_3,t_2+\aw+s_b;
2\aw+\sib+s_a)\right|_{w_3=0}\Bigr\} .
\end{align}
The evaluation of this term is much easier than that of the first
integral explained above since there is only one Schwinger parameter
$t_2$ for $1/\Ke$.  We have only to Laurent-expand the integrand with
respect to $\veps$ after making the change of integration variables
from $t_2$ to $u=\veps t_2$, and carry out the
$(t_1,u)$-integrations. After all, the whole of $E_{A,0}^{(a,b)}$
\eqref{eq:E_11-0_2b} is found to be given by
\begin{equation}
E_{A,0}^{(a,b)}
=O\bigl(\veps^{\min(2k^2-1,1)}\bigr)
+\left(s_a+\sib+2\aw\right)^{1-2k^2}\Bigl[
C_2(\sia)+\left(s_a+\sib+2\aw\right)C_1(\sia)\Bigr] ,
\label{eq:E_11-0_2b_final}
\end{equation}
where the first (second) term on the RHS corresponds to the first
(second) integral of \eqref{eq:E_11-0_2b}.

The first term on the RHS of \eqref{eq:E_11-0_2b_final},
$O\bigl(\veps^{\min(2k^2-1,1)}\bigr)$, vanishes in the limit
$\veps\to 0$ for $k^2>1/2$, while it is divergent for $k^2<1/2$.
Here, we {\em define} $E_{A,0}^{(a,b)}$ for a generic $k^2$ as
the ``analytic continuation'' from the region of sufficiently
large $k^2$ ($k^2>1/2$ in the present case). Thus, $E_{A,0}^{(a,b)}$
is simply given by the last term of \eqref{eq:E_11-0_2b_final}.
Eq.\ \eqref{eq:E_11-0_2b_final} has been obtained by keeping only the
first term of the Laurent expansion. The subleading term which has an
extra positive power $(\veps/u)^p$ contributes
$O\bigl(\veps^{\min(2k^2-1+p,1)}\bigr)$. This vanishes for $k^2>1/2$
and does not affect our definition of $E_{A,0}^{(a,b)}$ by analytic
continuation.
We apply this definition of $E_{A,0}^{(a,b)}$ by analytic
continuation from the region of sufficiently large $k^2$ also to other
$k^2$-dependent quantities; $E_{\ell,0}^{(a,b)}$ ($\ell=B,C$)
and $W_{ij}^{(a,b)}$.

The evaluation of $E_{B,0}^{(a,b)}$ and  $E_{C,0}^{(a,b)}$ is
similar except two points. First, they contain terms with {\em three}
$1/\Ke$. For such terms, we have to carry out the integration over the
three Schwinger parameters by considering $2^3-1=7$ regions
with at least one large parameters (see Appendix
\ref{app:SevenSubregions}).
Second, the obtained $E_{\ell,0}^{(a,b)}$ ($\ell=B,C$)
both contain $1/\veps$ terms, and, therefore, $\eveps$
\eqref{eq:eveps} multiplying them makes non-trivial contribution to
their $O(\veps^0)$ terms.
Then, we get the following results:\footnote{
If we adopt $\eaK$ instead of $\eaKe$ in the definition of
$\uu_i$ \eqref{eq:uu}, we have to replace all $1/\veps$ in
\eqref{eq:E_B0_2b_0-0} and \eqref{eq:E_C0_2b_0-0} with
$(1/\veps)+\pi$.
}
\begin{align}
E_{B,0}^{(a,b)}
&=\frac{1-k^2}{1-2k^2}\left(s_a+\sib+2\aw\right)^{1-2k^2}C_2(\sia)
\nn\\
&\quad
+\left(s_b+\sia+2\aw\right)^{1-2k^2}\left\{
-\frac12\,C_2(\sib)
+\left[\frac{1}{1-2k^2}\frac{3}{\pi^2}\invveps
-\left(s_b+\sia+2\aw\right)
\right]C_1(\sib)\right\} ,
\label{eq:E_B0_2b_0-0}
\\
E_{C,0}^{(a,b)}
&=-\frac{1}{2(1-2k^2)}
\left(s_b+\sia+2\aw\right)^{1-2k^2}\left[
2\,C_1(\sib)+C_2(\sib)\right]
\nn\\
&\quad
-\frac{1}{1-2k^2}
\left(\frac{3}{\pi^2}\invveps+1\right)
\left(s_a+\sib+2\aw\right)^{1-2k^2}C_1(\sia) .
\label{eq:E_C0_2b_0-0}
\end{align}
In particular, $E_{A,0}^{(a,b)}-E_{B,0}^{(a,b)}+E_{C,0}^{(a,b)}$,
which is related to the EOM against
$\cR{u_0^{(a)}}{u_{1A}^{(b)}-u_{1B}^{(b)}}$, is given by
\begin{align}
E_{A,0}^{(a,b)}-E_{B,0}^{(a,b)}+E_{C,0}^{(a,b)}
&=-\left\{\frac{k^2}{1-2k^2}\,C_2(\sia)
+\left[\frac{1}{1-2k^2}
\left(\frac{3}{\pi^2}\frac{1}{\veps}+1\right)-\sapib
\right]C_1(\sia)\right\}\sapib^{1-2k^2}
\nn\\
&\quad
-\left\{\frac{k^2}{1-2k^2}\,C_2(\sib)
+\left[\frac{1}{1-2k^2}
\left(\frac{3}{\pi^2}\frac{1}{\veps}+1\right)-\sbpia
\right]C_1(\sib)\right\}\sbpia^{1-2k^2} ,
\label{eq:E_11,0-E_12,0+E_13,0_2b}
\end{align}
where $\sapib$ and $\sbpia$ are defined by
\begin{equation}
\sapib=s_a+\sib+2\aw,\qquad \sbpia=s_b+\sia+2\aw .
\label{eq:s_a+ib}
\end{equation}
Eq.\ \eqref{eq:E_11,0-E_12,0+E_13,0_2b} implies that, in order for the
EOM against $\cR{u_0^{(a)}}{u_{1A}^{(b)}-u_{1B}^{(b)}}$ to hold for
any $k^2$, $v_{1/R}^{(a)}(\sia)$ and $v_{1/R}^{(b)}(\sib)$ must
be such that satisfy
\begin{equation}
\int_0^\infty\!d\si\,v_{1/R}(\si)\Pmatrix{
C_1(\si) \\ C_2(\si)}=0 .
\label{eq:cond_v_1/R}
\end{equation}
In this case, the EOMs against
$\cR{u_0^{(a)}}{u_{1A}^{(b)}}$
and
$\cR{u_0^{(a)}}{u_{1B}^{(b)}}$
hold for any $\xi$.
Furthermore, the EOM against $u_{1A/B}$ also holds for any $\xi$ as
seen from \eqref{eq:E_1ell_2b_1-1}.

The condition \eqref{eq:cond_v_1/R} restricts the first few terms of
the series expansion of $1/R(\Ke)$ with respect to $\Ke$.
In fact, expanding the expression \eqref{eq:LandR_by_v} for $1/R(\Ke)$
in powers of $\Ke$ and using the condition \eqref{eq:cond_v_1/R},
we obtain
\begin{equation}
\frac{1}{R(\Ke)}=\invKe\left\{1+(\aw+1)\Ke
+\frac12\,\aw\left(\aw+2\right)\Ke^2+O\bigl(\Ke^3\bigr)\right\} .
\label{eq:1/R_C1,2}
\end{equation}

\subsubsection{$(m_a,n_a)=(m_b,n_b)=(0,0)$}

The complete evaluation of $E_{\ell,0}^{(a,b)}$ for
$(m_a,n_a)=(m_b,n_b)=(0,0)$ is much harder than that for $(1,1)$.
Here, however, we need only their $1/\veps^2$ part:
\begin{align}
\Pmatrix{E_{A,0}^{(a,b)}\\[2mm] E_{B,0}^{(a,b)}}
&=-\Pmatrix{1 \\[2mm] 2}\frac{3}{\pi^2}\frac{1}{1-2k^2}
\left(s_b+\sia+2\aw\right)^{1-2k^2}\times\frac{1}{\veps^2}
+O\!\left(\frac{1}{\veps}\right) ,
\nn\\
E_{C,0}^{(a,b)}&=\frac{3}{\pi^2}\frac{1}{1-2k^2}
\left(s_a+\sib+2\aw\right)^{1-2k^2}\times\frac{1}{\veps^2}
+O\!\left(\frac{1}{\veps}\right) .
\end{align}
This result implies that the $1/\veps^2$ part of the EOM against
$\cR{u_{1A/B}^{(a)}}{u_0^{(b)}}$ cannot vanish for any choice of
$\xi_a$ (and, in particular, for $\xi_a=-1$ determined
in Sec.\ \ref{sec:E_1ell_2b_00}) at least in the case $(a)=(b)$
which we take in the end.
Therefore, we do not consider the case $(m_a,n_a)=(m_b,n_b)=(0,0)$
in the rest of this section.

\subsection{$\omega_{ij}^{(a,b)}$ for $(m_a,n_a)=(m_b,n_b)=(1,1)$}
\label{sec:omega_2b_1-1}

Let us complete the BV analysis around the 2-brane solution by
evaluating the matrix $\omega_{ij}^{(a,b)}$ \eqref{eq:omega_ij^(a,b)}
for $(m_a,n_a)=(m_b,n_b)=(1,1)$ (see
\eqref{eq:W_03}--\eqref{eq:W_1A2B^4} for explicit expressions of
$W_{i,j}^{(a,b)}$).
First, for $W_{0,3}^{(a,b)}$ \eqref{eq:W_03}, we obtain\footnote{
The first and the second terms in \eqref{eq:W_03} are of
$O\bigl(\veps^{\min(2k^2,1)}\bigr)$ and $O(\veps)$, respectively.
}
\begin{equation}
W_{0,3}^{(a,b)}=O\bigl(\veps^{\min(2k^2,1)}\bigr) .
\end{equation}
Namely, $W_{0,3}^{(a,b)}$ defined by analytic continuation is
identically equal to zero.
Next, $W_{1A,2A}^{(a,b)(1)}$ \eqref{eq:W_1A2A^1} and
$W_{1A,2A}^{(a,b)(2)}$ \eqref{eq:W_1A2A^2} constituting
$W_{1A,2A}^{(a,b)}$ by \eqref{eq:W_1A2A=W_1A2A^1+W_1A2A^2} are given
by
\begin{align}
W_{1A,2A}^{(a,b)(1)}&=\int\!c\,V_{-k}\,e^{-(2\aw+s_b+\sia)K}
cKc\,V_k\,e^{-(2\aw+s_a+\sib)K}
=f(s_a,s_b,\sia,\sib) ,
\label{eq:W_1A2A^1_2b}
\\[2mm]
W_{1A,2A}^{(a,b)(2)}&=O\bigl(\veps^{\min(2k^2,1)}\bigr) ,
\end{align}
with
\begin{equation}
f(s_a,s_b,\sia,\sib)
=\left[\left(1+\frac{s_a+s_b+\sia+\sib}{\pi}\right)
\sin\left(\frac{\frac{\pi}{2}+s_b+\sia}{
1+\frac{1}{\pi}\left(s_a+s_b+\sia+\sib\right)}\right)
\right]^{2(1-k^2)} .
\label{eq:f}
\end{equation}
Therefore, $\omega_{1A,2A}^{(a,b)}$ defined by analytic continuation
is independent of $\xi_a$ and is given by (see \eqref{eq:def_W_ij})
\begin{equation}
\omega_{1A,2A}^{(a,b)}=\intfour\!\! f(s_a,s_b,\sia,\sib) .
\label{eq:omega_1A2A_by_f}
\end{equation}
Finally, for $W_{1A,2B}^{(a,b)}$ given by
\eqref{eq:W_1A2B=sumW_1A2A^k}, we obtain the following result after
the analytic continuation:
\begin{align}
W_{1A,2B}^{(a,b)(1)}&=0 ,
\nn\\
W_{1A,2B}^{(a,b)(2)}&=\frac12 \sapib^{1-2k^2}
\left[C_2(\sia)+\sapib\,C_1(\sia)\right] ,
\nn\\
W_{1A,2B}^{(a,b)(3)}&=-\frac12 \sbpia^{1-2k^2}
\left[C_2(\sib)+\sbpia\,C_1(\sib)\right] ,
\nn\\
W_{1A,2B}^{(a,b)(4)}
&=\frac12\frac{1-k^2}{1-2k^2}\left[
\sapib^{1-2k^2}C_2(\sia)-\sbpia^{1-2k^2}C_2(\sib)\right]
-\left(\frac{1}{1-2k^2}\frac{3}{\pi^2}\frac{1}{\veps}
-\sapib\right)\sapib^{1-2k^2}C_1(\sia) .
\label{eq:W_1A2B^1,2,3,4_2b}
\end{align}
Assuming that $v_{1/R}^{(a)}$ and $v_{1/R}^{(b)}$ both
satisfy the condition \eqref{eq:cond_v_1/R}, our result
\eqref{eq:W_1A2B^1,2,3,4_2b} implies that $\omega_{1A,2B}^{(a,b)}$ is
equal to zero for any $(\xi_a,\xi_b)$.

Summarizing, we have obtained
\begin{equation}
\omega_{0,3}^{(a,b)}=0,
\qquad
\omega_{1A,2A}^{(a,b)}=1+O\!\left(k^2-1\right),
\qquad
\omega_{1A,2B}^{(a,b)}=0 ,
\label{eq:omega_2b}
\end{equation}
and, from \eqref{eq:Omega},
\begin{equation}
\Pmatrix{
\omega_{1A,2A}^{(a,b)} &\omega_{1A,2B}^{(a,b)}
\\[2mm]
\omega_{1B,2A}^{(a,b)} &\omega_{1B,2B}^{(a,b)}
}
=\Pmatrix{1 & 0 \\[1mm] 1 & 0}\omega_{1A,2A}^{(a,b)} .
\label{eq:Omega_2b}
\end{equation}

\subsection{The action of the fluctuation with $(m,n)=(1,1)$}
\label{sec:action_2b}

The above result, in particular, $\omega_{0,3}^{(a,b)}=0$, implies
that the present $\omega_{ij}^{(a,b)}$ for the six BV-states is
degenerate. From $\omega_{0,3}^{(a,b)}=0$ and \eqref{eq:Omega_2b}, we
see that the rank of the $6\times 6$ matrix $\omega_{ij}^{(a,b)}$ is
two, and that there exists effectively the following four
equivalences:\footnote{
For a state $w$, $w\sim 0$ implies that
$\omega_{w,j}^{(a,b)}=\int\!w^{(a)}u_j^{(b)}$
vanishes for any $u_j^{(b)}$ in the six BV states.
Note that $\calQ w\sim0$ follows from $w\sim 0$ due to the property
\eqref{eq:partint_calQ}.
}
\begin{equation}
u_0^{(a)}\sim 0,\qquad
u_{2B}^{(a)}\sim 0,\qquad
u_3^{(a)}\sim 0,\qquad
u_{1A}^{(a)}\sim u_{1B}^{(a)} .
\label{eq:eqrel}
\end{equation}
Therefore, we express the fluctuation $\Phi$ around the solution
$\calPe$ in terms of only $u_{1A}$ and $u_{2A}$ which are non-trivial
and independent:
\begin{equation}
\Phi=\int_k\left(u_{1A}(k)\,\chi(k)+u_{2A}(k)\,\chi_\star(k)\right) .
\label{eq:expandPhi_2b}
\end{equation}
Here, we have chosen as $v_{1/R}(\si)$ defining $u_i$ a suitable one
satisfying the condition \eqref{eq:cond_v_1/R}, and omitted the
superscript $(a)$ as in Sec.\ \ref{sec:calS0_nondegen}.
Plugging \eqref{eq:expandPhi_2b} into the kinetic term
$\calS_0[\Phi]$ \eqref{eq:calS_0} and using \eqref{eq:int_u1calQu1},
we obtain
\begin{equation}
\calS_0[\Phi]=-\int_k\!\frac12\,\omega_{1A,2A}(k)
\left(k^2-1\right)\chi(-k)\,\chi(k) .
\label{eq:calS0_2b}
\end{equation}
Since we have $\omega_{1A,2A}(k^2=1)=1$, the expansion
\eqref{eq:expandPhi_2b} and the action \eqref{eq:calS0_2b} are
essentially the same as \eqref{eq:Psi=sum_u_phi_tach_pert} and
\eqref{eq:S0_tach_pert}, respectively, for the tachyon field on the
unstable vacuum. The present $\chi$ represents a physical tachyon
field.

Finally, let us interpret the above result in the context of the BRST
cohomology problem.
Using the truncation \eqref{eq:eqrel} and discarding the EOM terms in
the BRST transformation formula \eqref{eq:calQu_i}, we obtain the
following equations for the remaining $u_{1A}$ and $u_{2A}$:
\begin{equation}
\calQ\,u_{1A}=\left(1-k^2\right)u_{2A},
\qquad
i\calQ\,u_{2A}=0 .
\end{equation}
The first equation and the fact that $u_0\sim 0$, namely, that there
is no candidate BRST parent of $u_{1A}$, imply that $u_{1A}$
at $k^2=1$ is a physical state belonging to
$\mathrm{Ker}\calQ/\mathrm{Im}\calQ$.

\section{BV analysis around the tachyon vacuum solution}
\label{sec:BV_TV}

In this section, we repeat the BV analysis of the previous section by
taking $G_\textrm{tv}$ \eqref{eq:twoG} which represents the tachyon
vacuum. We expect of course that the matrix $\omega_{ij}$ of the six
BV states $u_i$ is non-degenerate and therefore the excitations they
describe are unphysical ones.
As $(m,n)$ for $(L,R)$, we consider here again only the two cases,
$(1,1)$ and $(0,0)$.

\subsection{EOM against $u_{1A/B}$ and $\cR{u_{1A/B}^{(a)}}{u_0^{(b)}}$}

First, $E_{\ell}$ ($\ell=A,B,C$) \eqref{eq:def_E_1ell}
for the EOM against $u_{1A/B}$ are calculated to be given by
\begin{equation}
\EA=\EB=2,\quad \EC=0
\quad\textrm{for}\quad (m,n)=(1,1) ,
\label{eq:E1ell_1-1_tv}
\end{equation}
and
\begin{equation}
\EA=\EB=\EC=0
\quad\textrm{for}\quad (m,n)=(0,0) .
\label{eq:E1ell_0-0_tv}
\end{equation}
The result \eqref{eq:E1ell_1-1_tv} implies that the EOMs against
$u_{1A}$ and $u_{1B}$ cannot be satisfied for any $\xi$
in the case $(m,n)=(1,1)$.
On the other hand, EOMs against $u_{1A/B}$ both hold for an arbitrary
$\xi$ in the case $(0,0)$.
Therefore, in the rest of this section, we consider only the latter
case $(m,n)=(0,0)$.

Next, $E_{\ell,0}^{(a,b)}$ ($\ell=A,B,C$) \eqref{eq:def_E_1ell-0} for
the EOM against $\cR{u_{1A/B}^{(a)}}{u_0^{(b)}}$ in the case
$(m,n)=(0,0)$ are found to be given by
\begin{equation}
E_{A,0}^{(a,b)}=O\bigl(\veps^{\min(2k^2,1)}\bigr),
\quad
E_{B,0}^{(a,b)}=O\bigl(\veps^{\min(2k^2-1,1)}\bigr),
\quad
E_{C,0}^{(a,b)}=O\bigl(\veps^{\min(2k^2-1,1)}\bigr) .
\end{equation}
Namely, $E_{\ell,0}^{(a,b)}$ defined by analytic continuation are all
equal to zero.

Summarizing, all the EOM tests are satisfied for any
$\left(v_L,v_{1/R},\xi\right)$
in the case $(m,n)=(0,0)$.
The choice $(m,n)=(1,1)$ is excluded by the EOM test against
$u_{1A/B}$.

\subsection{$\omega_{ij}^{(a,b)}$ for $(m_a,n_a)=(m_b,n_b)=(0,0)$
and $\xi_a=\xi_b=0$}
\label{sec:omega_tv}

For the tachyon vacuum solution and for $(m_a,n_a)=(m_b,n_b)=(0,0)$,
we find that $W_{0,3}^{(a,b)}$ \eqref{eq:W_03} is non-trivial and is
given by
\begin{equation}
W_{0,3}^{(a,b)}=-\left(s_b+\sia+2\aw\right)^{2(1-k^2)}
+O\bigl(\veps^{\min(2k^2,1)}\bigr) ,
\label{eq:W_03_tv}
\end{equation}
where the last term should be discarded by analytic continuation.

Next, for $\omega_{1A,2A}^{(a,b)}$ and $\omega_{1A,2B}^{(a,b)}$, we
have to specify $(\xi_a,\xi_b)$. Here, we consider the simplest
case of $(\xi_a,\xi_b)=(0,0)$, for which we need to calculate
only $W_{1A,2A}^{(a,b)(1)}$ \eqref{eq:W_1A2A^1} and
$W_{1A,2B}^{(a,b)(1)}$ \eqref{eq:W_1A2B^1}.
We see that the former, which is independent of $\Ge$ and depends on
$(m_a,n_a)$ and $(m_b,n_b)$ only through the differences $m_b-n_a$ and
$m_a-n_b$, is the same as \eqref{eq:W_1A2A^1_2b} for the 2-brane
solution:
\begin{equation}
W_{1A,2A}^{(a,b)(1)}=f(s_a,s_b,\sia,\sib) ,
\end{equation}
with $f$ given by \eqref{eq:f}. As for the latter, we find that
\begin{equation}
W_{1A,2B}^{(a,b)(1)}=O(\veps) .
\end{equation}
Our result implies that
\begin{equation}
\omega_{0,3}^{(a,b)}=-1+O(k^2-1),
\qquad
\omega_{1A,2A}^{(a,b)}=1+O(k^2-1),
\qquad
\omega_{1A,2B}^{(a,b)}=0 ,
\label{eq:omega_tv}
\end{equation}
for any $v_L^{(a/b)}$ and $v_{1/R}^{(a/b)}$.
Since $\omega_{ij}$ for the six BV states $u_i$ are non-degenerate,
the general argument of Sec.\ \ref{sec:calS0_nondegen} does apply to
the present case.
{}From \eqref{eq:omega_tv}, the function $\varpi(k)$ \eqref{eq:varpi}
is given by\footnote{\label{fn:negative_varpi}
This $\varpi(k)$ is not necessarily non-negative, and this may be a
problem for the hermiticity of $\wt{u}_i$ related to $u_i$ by
\eqref{eq:wtu} containing $\sqrt{\varpi(k)}$.
For example, in the simplest case of $L=R=1$,
we have $\omega_{1A,2A}(k)\equiv 1$ and hence $\varpi(k)$ is negative
for $k^2<1$.
Though the hermiticity of the original BV states $u_i$ itself is a
problem as we mentioned in Sec.\ \ref{sec:hermiticity}, one way to
resolve the negative $\varpi$ problem would be to Wick-rotate to the
Euclidean space-time where we have $k^2\ge 0$ (the negative $\varpi$
region, $0\le k^2<1$, should be regarded as an artifact of the
tachyon).
}
\begin{equation}
\varpi(k)=\left(k^2-1\right)\omega_{1A,2A}(k)
=k^2-1+O\!\left((k^2-1)^2\right) ,
\end{equation}
and the fluctuation around the tachyon vacuum we have constructed
is an unphysical one with $m^2=-1$.

Finally, our result is interpreted in the BRST cohomology problem as
follows. On the mass-shell $k^2=1$, $\omega_{ij}$ is reduced to
\begin{equation}
\omega_{0,3}=-1,
\qquad
\Pmatrix{\omega_{1A,2A} & \omega_{1A,2B}
\\[2mm]
\omega_{1B,2A} & \omega_{1B,2B}}
=\Pmatrix{1 & 0 \\ 0 & 0},
\qquad
\left(k^2=1\right).
\end{equation}
From this we find that $u_{1B}\sim 0$ and $u_{2B}\sim 0$ at $k^2=1$.
Then, from \eqref{eq:calQu_i}, we obtain the following BRST
transformation rule for the remaining $(u_0,u_{1A},u_{2A},u_3)$:
\begin{equation}
i\calQ u_0=u_{1A},
\qquad
\calQ u_{1A}=0,
\qquad
i\calQ u_{2A}=u_3,
\qquad
\calQ u_3=0,
\qquad \left(k^2=1\right).
\end{equation}
This implies, in particular, that the candidate physical state
$u_{1A}$ is a trivial element of
$\mathrm{Ker}\calQ/\mathrm{Im}\calQ$.
Of course, this cannot be a proof of the total absence of physical
excitations with $m^2=-1$ on the tachyon vacuum.

\section{Summary and discussions}
\label{sec:summary}

In this paper, we carried out the analysis of the six tachyon BV
states for the 2-brane solution and for the tachyon vacuum solution in
CSFT. This set of six states was chosen from the requirement that the
EOM of the solution holds against the states and their commutators.
We found that the matrix $\omega_{ij}$ defining the BV equation
is degenerate and therefore the tachyon mode is physical for the
2-brane solution.
On the other hand, $\omega_{ij}$ is non-degenerate on the
tachyon vacuum solution, implying that the candidate tachyon field is
in fact unphysical there. These results are in agreement with our
expectation and the general proof of the non-existence of
physical excitations on the tachyon vacuum \cite{EllSch}.

Our analysis in this paper is incomplete in several respects.
First, we have not identified all of the four tachyon fields of the
same $m^2=-1$ which should exist on the 2-brane solution.
Secondly and more importantly, we must resolve the problem that our
six tachyon BV states \eqref{eq:sixstates_MB} do not
satisfy the hermiticity condition \eqref{eq:Herm_u_i}.
Even if we put aside this problem, there are a number of questions
to be understood concerning our tachyon BV states:
\begin{itemize}
\item
The construction of our tachyon BV states \eqref{eq:sixstates_MB}
is not a unique one. In particular, the division of $i\calQ u_0$ into
$u_{1A}$ and $u_{1B}$ and that of $\calQ u_{1A/B}$ into $u_{2A}$ and
$u_{2B}$ (see \eqref{eq:calQu_i}) have much arbitrariness which is not
reduced to a linear recombination among the two states.
We have to confirm that the (non-)existence of physical tachyon
fluctuation does not depend on the choice of the tachyon BV states so
long as they satisfy the EOM conditions.
(Or we have to establish a criterion for selecting a particular set of
the BV states besides the EOM conditions.)

In this paper, we introduced one parameter $\xi$ representing an
arbitrariness of the tachyon BV states (recall
\eqref{eq:sixstates_MB}).
For the 2-brane solution and for $(m,n)=(1,1)$ and $v_{1/R}$
satisfying \eqref{eq:cond_v_1/R} from the EOM conditions,
we found in Sec.\ \ref{sec:omega_2b_1-1} that the matrix
$\omega_{ij}(k)$ is totally independent of the parameter $\xi$,
implying that a physical tachyon fluctuation exists for any $\xi$.
For the tachyon vacuum solution and for $(m,n)=(0,0)$, the results
for $\omega_{0,3}$ given in \eqref{eq:W_03_tv} and
\eqref{eq:omega_tv} are independent of $\xi$.
Though we have to evaluate other $\omega_{ij}$ for confirming the
non-degeneracy of the $2\times 2$ part $\Omega$ \eqref{eq:Omega}
with $\det \Omega=\omega_{0,3}\varpi$, the fact that
$\omega_{0,3}(k)\ne 0$ supports that the present set of the tachyon BV
states is an unphysical one for any $\xi$.

Besides the analysis presented in Secs.\ \ref{sec:BV_2b} and
\ref{sec:BV_TV}, we carried out the analysis also for the BV states
\eqref{eq:sixstates_MB} using another choice of $\uu_i$ given by
\eqref{eq:another_uu}.
The results for this BV states are mostly the same as those for the BV
states using $\uu_i$ of \eqref{eq:uu}.
First, for the 2-brane solution, the EOM conditions are all satisfied
for $(m,n)=(1,1)$ and $v_{1/R}$ satisfying \eqref{eq:cond_v_1/R},
and we obtain, in the particular case of $\xi=1$,
\begin{equation}
\omega_{0,3}=0 ,
\qquad
\omega_{1A,2A}=-1+O\left(k^2-1\right) ,
\qquad
\omega_{1A,2B}=0 ,
\label{eq:omega_2b_another_uu}
\end{equation}
where $\omega_{1A,2A}$ is given by the following $W_{1A,2A}^{(a,b)}$:
\begin{equation}
W_{1A,2A}^{(a,b)}=f(s_a,s_b,\sia,\sib)
-\left(\frac{\pi}{2}+s_a+\sib\right)^{2(1-k^2)}
-\left(\frac{\pi}{2}+s_b+\sia\right)^{2(1-k^2)} ,
\end{equation}
with $f$ defined by \eqref{eq:f}.
This result should be compared with \eqref{eq:omega_2b} for the choice
\eqref{eq:uu} of $\uu_i$ adopted in Sec.\ \ref{sec:BV_2b}.
Eq.\ \eqref{eq:omega_2b_another_uu} implies that $\omega_{ij}$
is degenerate and the tachyon is physical.
However, the fact that $\omega_{1A,2A}=-1$ at $k^2=1$ implies that
the the tachyon field kinetic term given by \eqref{eq:calS0_2b} has
the wrong sign, namely, that the physical tachyon is a negative norm
one. Of course, we have to resolve the hermiticity problem before taking
this problem seriously.
Secondly, for the tachyon vacuum solution and for $(m,n)=(0,0)$ and
$\xi=1$, we found that $\omega_{ij}$ is non-degenerate and hence the
fluctuation is an unphysical one. The main difference from the case of
$\uu_i$ given by \eqref{eq:uu} is that $\omega_{ij}$ are of
$O(1/\veps)$; for example,
$\omega_{0,3}=-(\pi/2)^{2(1-k^2)}\left[
1+6/\left(\pi^3\left(2k^2-1\right)\veps\right)\right]$
for $L=R=1$.

These two results, one concerning the parameter $\xi$ and the other
for another choice \eqref{eq:another_uu} of $\uu_i$, may support the
expectation that the (un)physicalness of the tachyon fluctuation is
insensitive to the details of the choice of the BV states.
In any case, we need a deeper understanding and general proof of this 
expectation.

\item
We have restricted our analysis of the kinetic term $\calS_0$
\eqref{eq:calS_0} only to the six tachyon BV states and
ignored the presence of all other states.
For this analysis to be truly justified, we have to show that the
complete set of the BV states of fluctuation can be constructed by
adding to our set of tachyon BV states its complementary set of BV
states which are orthogonal (in the sense of $\omega_{ij}=0$) to the
former set.

\item
As $(m,n)$ specifying the leading small $\Ke$ behavior of $L(\Ke)$ and
$R(\Ke)$, we have considered only the two cases, $(1,1)$ and $(0,0)$.
We should examine whether there are other allowed $(m,n)$ passing the
EOM tests, and if so, we must clarify the relationship among the BV
bases with different $(m,n)$.

\end{itemize}
Finally, we have to extend our analysis to more generic $n$-brane
solutions (including the exotic one with $n=-1$), and also to
fluctuations other than the tachyon mode.

\section*{Acknowledgments}
This work  was supported in part by a Grant-in-Aid for
Scientific Research (C) No.~25400253 from JSPS.

\appendix

\section{$\KBc$ algebra and correlators}
\label{app:KBc}

Here, we summarize the $\KBc$ algebra and the correlators which we
used in the text.
The elements of the $\KBc$ algebra satisfy
\begin{equation}
\CR{B}{K}=0,\quad\left\{B,c\right\}=1,\quad
B^2=c^2=0,
\label{eq:KBcA}
\end{equation}
and
\begin{equation}
\QB B=K,\quad\QB K=0,\quad\QB c=cKc .
\end{equation}
Their ghost numbers are
\begin{equation}
\Ngh(K)=0,\quad \Ngh(B)=-1,\quad \Ngh(c)=1 .
\end{equation}

In the text and in Appendix \ref{app:EEW}, there appear the following
CSFT integrations:
\begin{align}
\int\!Bc\,e^{-t_1 K}c\,e^{-t_2 K}c\, e^{-t_3 K}c\,e^{-t_4 K}
&=\VEV{Bc(0)c(t_1)c(t_1+t_2)c(t_1+t_2+t_3)}_{t_1+t_2+t_3+t_4},
\label{eq:Bcccc_col}
\\[3mm]
\int\!c\,e^{-t_1 K}c\,e^{-t_2 K}c\,e^{-t_3 K}
&=\VEV{c(0)c(t_1)c(t_1+t_2)}_{t_1+t_2+t_3}.
\end{align}
They are given in terms of the correlators on the cylinder with
infinite length and the circumference $\ell$:
\begin{align}
\VEV{B\,c(z_1)c(z_2)c(z_3)c(z_4)}_{\ell}
&=\left(\frac{\ell}{\pi}\right)^2\biggl\{
-\frac{z_1}{\pi}\sin\!\left[\frac{\pi}{\ell}(z_2-z_3)\right]
\sin\!\left[\frac{\pi}{\ell}(z_2-z_4)\right]
\sin\!\left[\frac{\pi}{\ell}(z_3-z_4)\right]
\nn\\
&\qquad
+\frac{z_2}{\pi}\sin\!\left[\frac{\pi}{\ell}(z_1-z_3)\right]
\sin\!\left[\frac{\pi}{\ell}(z_1-z_4)\right]
\sin\!\left[\frac{\pi}{\ell}(z_3-z_4)\right]
\nn\\
&\qquad
-\frac{z_3}{\pi}\sin\!\left[\frac{\pi}{\ell}(z_1-z_2)\right]
\sin\!\left[\frac{\pi}{\ell}(z_1-z_4)\right]
\sin\!\left[\frac{\pi}{\ell}(z_2-z_4)\right]
\nn\\
&\qquad
+\frac{z_4}{\pi}\sin\!\left[\frac{\pi}{\ell}(z_1-z_2)\right]
\sin\!\left[\frac{\pi}{\ell}(z_1-z_3)\right]
\sin\!\left[\frac{\pi}{\ell}(z_2-z_3)\right]\biggr\} ,
\label{eq:VEVBcccc}
\\[3mm]
\VEV{c(z_1)c(z_2)c(z_3)}_{\ell}&=\left(\frac{\ell}{\pi}\right)^3
\sin\!\left[\frac{\pi}{\ell}(z_1-z_2)\right]
\sin\!\left[\frac{\pi}{\ell}(z_1-z_3)\right]
\sin\!\left[\frac{\pi}{\ell}(z_2-z_3)\right] .
\label{eq:VEVccc}
\end{align}
Finally, the matter correlator is given by
\begin{equation}
\bigl\langle e^{ik\cdot X(z,\ol{z})}
\,e^{ik'\cdot X(z',\ol{z}')}\bigr\rangle_\ell^\textrm{matt}
=\abs{\frac{\ell}{\pi}\sin\frac{\pi(z-z')}{\ell}}^{-2k^2}
\times(2\pi)^{26}\delta^{26}(k+k') .
\label{eq:matt_correlator}
\end{equation}

\section{$E_{\ell}$, $E_{\ell,0}^{(a,b)}$,
$W_{0,3}^{(a,b)}$, $W_{1A,2A}^{(a,b)}$ and $W_{1A,2B}^{(a,b)}$}
\label{app:EEW}

In this appendix, we present explicit expressions of the quantities
defined by \eqref{eq:def_E_1ell}, \eqref{eq:def_E_1ell-0} and
\eqref{eq:def_W_ij}.
In these expressions, $\cCR{\Ke}{\Ge}$ denotes the following
abbreviation:
\begin{equation}
\cCR{\Ke}{\Ge}=\Ke\,c\,\Ge-\Ge\,c\,\Ke .
\label{eq:cCRKeGe}
\end{equation}

\noindent
\underline{$E_{\ell}(s,\si)$ ($\ell=A,B,C$)}
\begin{align}
\EA&=\veps\int\!Bc\left(1-\Ge\right)\Ke^m e^{-(\aw+s)\Ke}
c\,\frac{e^{-(\aw+\si)\Ke}}{\Ge\Ke^n}\,c\cCR{\Ke}{\Ge} ,
\label{eq:E_11}
\\
\EB&=\left(1-k^2\right)\veps\int\!
BcKc\,\frac{e^{-(\aw+\si)\Ke}}{\Ge\Ke^n}
\,c\cCR{\Ke}{\Ge}\left(1-\Ge\right)\frac{e^{-(\aw+s)\Ke}}{\Ke^{1-m}}
\nn\\
&\qquad
+\veps^2\int\!Bc\left(1-\Ge\right)\frac{e^{-(\aw+s)\Ke}}{\Ke^{1-m}}
\,c\,\frac{e^{-(\aw+\si)\Ke}}{\Ge\Ke^n}\,c\cCR{\Ke}{\Ge},
\quad(k^2=0) ,
\label{eq:E_12}
\\
\EC&=-\veps^2\int\!Bc\,\frac{e^{-(\aw+\si)\Ke}}{\Ge\Ke^n}
c\cCR{\Ke}{\Ge}c\left(1-\Ge\right)\Ke^{m-1}\,e^{-(\aw+s)\Ke} .
\label{eq:E_13}
\end{align}
In \eqref{eq:E_12}, we have kept $k^2$ to make explicit the origin of
the term, though we of course have to put $k_\mu=0$ due to
momentum conservation.
We have omitted the space-time volume $(2\pi)^{26}\delta^{26}(k=0)=VT$
on the RHS of \eqref{eq:E_11}--\eqref{eq:E_13}.

\noindent
\underline{$E_{\ell,0}^{(a,b)}(s_a,s_b,\sia,\sib)$
($\ell=A,B,C$)}
\begin{align}
E_{A,0}^{(a,b)}&=-\veps\int\!Bc\,V_k
\frac{e^{-(\aw+\sib)\Ke}}{\Ge\Ke^{n_b}}c\cCR{\Ke}{\Ge}
\CR{c}{\left(1-\Ge\right)\Ke^{m_a}\,e^{-(\aw+s_a)\Ke}}V_{-k}
\frac{e^{-(2\aw+s_b+\sia)\Ke}}{\Ke^{1+n_a-m_b}}
\nn\\
&\quad
-\veps\int\!Bc\,V_k\frac{e^{-(2\aw+s_a+\sib)\Ke}}{\Ke^{n_b-m_a}}
\,c\,V_{-k}\frac{e^{-(\aw+\sia)\Ke}}{\Ge\Ke^{n_a}}
\,c\cCR{\Ke}{\Ge}\left(1-\Ge\right)
\frac{e^{-(\aw+s_b)\Ke}}{\Ke^{1-m_b}} ,
\label{eq:E_11_0}
\\
E_{B,0}^{(a,b)}
&=-\left(1-k^2\right)\veps\!\int\!BcKc\,V_{-k}
\frac{e^{-(2\aw+s_b+\sia)\Ke}}{\Ke^{1+n_a-m_b}}V_k
\CR{c}{\frac{e^{-(\aw+\sib)\Ke}}{\Ge\Ke^{n_b}}}
\cCR{\Ke}{\Ge}\left(1-\Ge\right)\frac{e^{-(\aw+s_a)\Ke}}{\Ke^{1-m_a}}
\nn\\
&\quad
-\left(1-k^2\right)\veps\!\int\!
BcKc\,V_{-k}\frac{e^{-(\aw+\sia)\Ke}}{\Ge\Ke^{n_a}}c\cCR{\Ke}{\Ge}
\left(1-\Ge\right)\frac{e^{-(\aw+s_b)\Ke}}{\Ke^{1-m_b}}V_k
\frac{e^{-(2\aw+s_a+\sib)\Ke}}{\Ke^{1+n_b-m_a}}
\nn\\
&\quad
-\veps^2\int\!Bc\,V_k\frac{e^{-(\aw+\sib)\Ke}}{\Ge\Ke^{n_b}}
c\cCR{\Ke}{\Ge}
\CR{c}{\left(1-\Ge\right)\frac{e^{-(\aw+s_a)\Ke}}{\Ke^{1-m_a}}}
V_{-k}\frac{e^{-(2\aw+s_b+\sia)\Ke}}{\Ke^{1+n_a-m_b}}
\nn\\
&\quad
-\veps^2\int\!Bc\,V_k\frac{e^{-(2\aw+s_a+\sib)\Ke}}{\Ke^{1+n_b-m_a}}
c\,V_{-k}\frac{e^{-(\aw+\sia)\Ke}}{\Ge\Ke^{n_a}}c\cCR{\Ke}{\Ge}
\left(1-\Ge\right)\frac{e^{-(\aw+s_b)\Ke}}{\Ke^{1-m_b}} ,
\label{eq:E_12_0}
\\
E_{C,0}^{(a,b)}&=-\veps^2\!\int\!Bc\,V_k
\frac{ e^{-(\aw+\sib)\Ke}}{\Ge\Ke^{n_b}}c\cCR{\Ke}{\Ge}
\CR{c}{\frac{1}{\Ge}-1} \Ge\left(1-\Ge\right)
\frac{e^{-(\aw+s_a)\Ke}}{\Ke^{1-m_a}}V_{-k}
\frac{e^{-(2\aw+s_b+\sia)\Ke}}{\Ke^{1+n_a-m_b}}
\nn\\
&\quad
-\veps^2\int\!Bc\,V_{-k}\frac{e^{-(\aw+\sia)\Ke}}{\Ge\Ke^{n_a}}
\,c\,\Ge c\,\frac{\Ke}{\Ge}\,c\,\Ge
\left(1-\Ge\right)\frac{e^{-(\aw+s_b)\Ke}}{\Ke^{1-m_b}}
V_k\frac{e^{-(2\aw+s_a+\sib)\Ke}}{\Ke^{1+n_b-m_a}}
\nn\\
&\quad
+\veps^2\int\!Bc\,V_{-k}\frac{e^{-(\aw+\sia)\Ke}}{\Ge\Ke^{n_a}}
\,c\,\Ge\left(1-\Ge\right)\frac{e^{-(\aw+s_b)\Ke}}{\Ke^{1-m_b}}V_k
\CR{c}{\frac{e^{-(\aw+\sib)\Ke}}{\Ge\Ke^{n_b}}}
\nn\\
&\qquad\qquad\times
\cCR{\Ke}{\Ge}\left(1-\Ge\right)\frac{e^{-(\aw+s_a)\Ke}}{\Ke^{1-m_a}}
+\left[\mbox{the last term with }(a)\rightleftarrows (b)\right] .
\label{eq:E_13_0}
\end{align}

\noindent
\underline{$W_{0,3}^{(a,b)}(s_a,s_b,\sia,\sib)$}
\begin{align}
W_{0,3}^{(a,b)}
&=-\veps\int\!Bc\left(1-\Ge\right)\Ke^{m_b}e^{-(\aw+s_b)\Ke}
cKc\,V_k\frac{e^{-(2\aw+s_a+\sib)\Ke}}{\Ke^{1+n_b-m_a}}V_{-k}
\CR{c}{\frac{e^{-(\aw+\sia)\Ke}}{\Ge\Ke^{n_a}}}\Ge
\nn\\
&\quad
-\veps\int\!Bc\,V_{-k}\Ke^{m_b-n_a}e^{-(2\aw+s_b+\sia)\Ke}
cKc\,V_k\frac{e^{-(\aw+\sib)\Ke}}{\Ge\Ke^{n_b}}
\,c\,\Ge\left(1-\Ge\right)\frac{e^{-(\aw+s_a)\Ke}}{\Ke^{1-m_a}} .
\label{eq:W_03}
\end{align}

\noindent
\underline{$W_{1A,2A}^{(a,b)}(s_a,s_b,\sia,\sib)$}
\begin{equation}
W_{1A,2A}^{(a,b)}=W_{1A,2A}^{(a,b)(1)}+\xi_a W_{1A,2A}^{(a,b)(2)} ,
\label{eq:W_1A2A=W_1A2A^1+W_1A2A^2}
\end{equation}
with
\begin{align}
W_{1A,2A}^{(a,b)(1)}
&=\int\!c\,V_{-k}\Ke^{m_b-n_a}\,e^{-(2\aw+\sia+s_b)\Ke}
cKc\,V_k\Ke^{m_a-n_b}\,e^{-(2\aw+\sib+s_a)\Ke} ,
\label{eq:W_1A2A^1}
\\
W_{1A,2A}^{(a,b)(2)}&=\veps\int\!Bc\,V_{-k}\Ke^{m_b-n_a}
e^{-(2\aw+s_b+\sia)\Ke}cKc\,V_k\frac{e^{-(\aw+\sib)\Ke}}{\Ge\Ke^{n_b}}
\,c\,\Ge\left(1-\Ge\right)\frac{e^{-(\aw+s_a)\Ke}}{\Ke^{1-m_a}}
\nn\\
&\quad
+\veps\int\!Bc\left(1-\Ge\right)\Ke^{m_b}e^{-(\aw+s_b)\Ke}cKc
\,V_k\frac{e^{-(2\aw+s_a+\sib)\Ke}}{\Ke^{1+n_b-m_a}}V_{-k}
\CR{c}{\frac{e^{-(\aw+\sia)\Ke}}{\Ge\Ke^{n_a}}}\Ge .
\label{eq:W_1A2A^2}
\end{align}

\noindent
\underline{$W_{1A,2B}^{(a,b)}(s_a,s_b,\sia,\sib)$}
\begin{equation}
W_{1A,2B}^{(a,b)}
=\left(1-\xi_b\right)W_{1A,2B}^{(a,b)(1)}
+\xi_b\,W_{1A,2B}^{(a,b)(2)}
+\xi_a\left(1-\xi_b\right)W_{1A,2B}^{(a,b)(3)}
+\xi_a\xi_b\,W_{1A,2B}^{(a,b)(4)} ,
\label{eq:W_1A2B=sumW_1A2A^k}
\end{equation}
with
\begin{align}
W_{1A,2B}^{(a,b)(1)}&=\veps\!\!\int\! Bc\left(1-\Ge\right)\Ke^{m_a}
e^{-(\aw+s_a)\Ke}c\,V_{-k}\frac{e^{-(2\aw+s_b+\sia)\Ke}}{\Ke^{n_a-m_b}}
\,c\,V_k\frac{e^{-(\aw+\sib)\Ke}}{\Ge\Ke^{n_b}}\,c\,\Ge
+\left[(a)\rightleftarrows (b)\right] ,
\label{eq:W_1A2B^1}
\\
W_{1A,2B}^{(a,b)(2)}&=\left(1-k^2\right)\biggl\{
\veps\int\!BcKc\,V_k\frac{e^{-(2\aw+s_a+\sib)\Ke}}{\Ke^{n_b-m_a}}
c\,V_{-k}\frac{e^{-(\aw+\sia)\Ke}}{\Ge\Ke^{n_a}}
\,c\,\Ge\left(1-\Ge\right)\frac{e^{-(\aw+s_b)\Ke}}{\Ke^{1-m_b}}
\nn\\
&\quad
+\veps\int\!BcKc\,V_k
\frac{e^{-(\aw+\sib)\Ke}}{\Ge\Ke^{n_b}}\,c\,\Ge
\CR{\left(1-\Ge\right)\Ke^{m_a}e^{-(\aw+s_a)\Ke}}{c}
V_{-k}\frac{e^{-(2\aw+s_b+\sia)\Ke}}{\Ke^{1+n_a-m_b}}
\biggr\}
\nn\\
&\quad
+\veps^2\int\!Bc\left(1-\Ge\right)\Ke^{m_a}e^{-(\aw+s_a)\Ke}
c\,V_{-k}\frac{e^{-(2\aw+s_b+\sia)\Ke}}{\Ke^{1+n_a-m_b}}
c\,V_k\frac{e^{-(\aw+\sib)\Ke}}{\Ge\Ke^{n_b}}\,c\,\Ge
\nn\\
&\quad
+\veps^2\int\!Bc\left(1-\Ge\right)\frac{e^{-(\aw+s_b)\Ke}}{\Ke^{1-m_b}}
c\,V_k\frac{e^{-(2\aw+s_a+\sib)\Ke}}{\Ke^{n_b-m_a}}
c\,V_{-k}\frac{e^{-(\aw+\sia)\Ke}}{\Ge\Ke^{n_a}}\,c\,\Ge ,
\label{eq:W_1A2B^2}
\\
W_{1A,2B}^{(a,b)(3)}
&=-\veps^2\int\!Bc\,V_{-k}\frac{e^{-(2\aw+s_b+\sia)\Ke}}{\Ke^{n_a-m_b}}
c\,V_k\frac{e^{-(\aw+\sib)\Ke}}{\Ge\Ke^{n_b}}
c\,\Ge c\left(1-\Ge\right)\frac{e^{-(\aw+s_a)\Ke}}{\Ke^{1-m_a}}
\nn\\
&\quad
+\veps^2\int\!Bc\,V_{-k}\frac{e^{-(\aw+\sia)\Ke}}{\Ge\Ke^{n_a}}
c\,\Ge c\left(1-\Ge\right)\Ke^{m_b}e^{-(\aw+s_b)\Ke}
c\,V_k\frac{e^{-(2\aw+s_a+\sib)\Ke}}{\Ke^{1+n_b-m_a}} ,
\label{eq:W_1A2B^3}
\\
W_{1A,2B}^{(a,b)(4)}
&=\left(1-k^2\right)\biggl\{
-\veps^2\int\!BcKc\,V_k\frac{e^{-(\aw+\sib)\Ke}}{\Ge\Ke^{n_b}}
c\,\Ge c\left(1-\Ge\right)\frac{e^{-(\aw+s_a)\Ke}}{\Ke^{1-m_a}}
V_{-k}\frac{e^{-(2\aw+s_b+\sia)\Ke}}{\Ke^{1+n_a-m_b}}
\nn\\
&\quad
+\veps^2\int\!Bc\,V_{-k}\frac{e^{-(\aw+\sia)\Ke}}{\Ge\Ke^{n_a}}
c\,\Ge c\left(1-\Ge\right)\frac{e^{-(\aw+s_b)\Ke}}{\Ke^{1-m_b}}
\CR{K}{c}V_k\frac{e^{-(2\aw+s_a+\sib)\Ke}}{\Ke^{1+n_b-m_a}}
\biggr\}
\nn\\
&\quad
-\veps^3\int\!Bc\,V_{-k}\frac{e^{-(2\aw+s_b+\sia)\Ke}}{\Ke^{1+n_a-m_b}}
c\,V_k\frac{e^{-(\aw+\sib)\Ke}}{\Ge\Ke^{n_b}}
c\,\Ge c\left(1-\Ge\right)\frac{e^{-(\aw+s_a)\Ke}}{\Ke^{1-m_a}}
\nn\\
&\quad
+\veps^3\int\!Bc\,V_{-k}\frac{e^{-(\aw+\sia)\Ke}}{\Ge\Ke^{n_a}}
c\,\Ge c\left(1-\Ge\right)\frac{e^{-(\aw+s_b)\Ke}}{\Ke^{1-m_b}}
c\,V_k\frac{e^{-(2\aw+s_a+\sib)\Ke}}{\Ke^{1+n_b-m_a}} .
\label{eq:W_1A2B^4}
\end{align}

\section{Seven integration regions for three $1/\Ke$}
\label{app:SevenSubregions}

In Sec.\ \ref{sec:E_1ell-0_2b_11-11}, we explained how to evaluate
correlators with two $1/\Ke$ by dividing the integration region of the
corresponding Schwinger parameters into three subregions
\eqref{eq:ThreeRegions}.
Here, we extend this to the case of three $1/\Ke$ with the
corresponding Schwinger parameters $(t_1,t_2,t_3)$.

First, we parametrize $(t_1,t_2,t_3)$ in terms of another set of
variables $(u,x,p)$ as
\begin{equation}
t_1=\uoe\,x,\qquad t_2=\uoe\left(1-x\right)p
\qquad t_3=\uoe\left(1-x\right)\left(1-p\right) ,
\end{equation}
which satisfies $t_1+t_2+t_3=u/\veps$.
The integration range of $(u,x,p)$ is $0\le u<\infty$,
$0\le x,p\le 1$.
For a correlator multiplied by a positive power of $\veps$,
we have only to consider the integration regions where at least one
of the three $t_i$ are large, and, in the present case, there are
seven such regions shown in Table \ref{tbl:SevenRegions}.
\begin{table}[h]
\centering
\begin{tabular}{|c|c|c|c|}
\hline
Region & $t_1$ & $t_2$ & $t_3$ \\
\hline\hline
IA & finite & finite & $\infty$ \\
\hline
IB & finite & $\infty$ & $\infty$ \\
\hline
IC & finite & $\infty$ & finite \\
\hline
IIA & $\infty$ & finite & $\infty$ \\
\hline
IIB & $\infty$ & $\infty$ & $\infty$ \\
\hline
IIC & $\infty$ & $\infty$ & finite \\
\hline
III & $\infty$ & finite & finite \\
\hline
\end{tabular}
\caption{Seven integration regions}
\label{tbl:SevenRegions}
\end{table}

In each of the seven regions, we adopt the following set of three
integration variables:
\begin{align}
\textrm{IA}:&\quad(u,y,z)\quad\textrm{with}\quad
x=\eou\,y,\quad p=\eou\,z ,
\nn\\
\textrm{IB}:&\quad(u,y,p)\quad\textrm{with}\quad
x=\eou\,y ,
\nn\\
\textrm{IC}:&\quad(u,y,z)\quad\textrm{with}\quad
x=\eou\,y,\quad 1-p=\eou\,z ,
\nn\\
\textrm{IIA}:&\quad(u,x,z)\quad\textrm{with}\quad
p=\eou\,z ,
\nn\\
\textrm{IIB}:&\quad(u,x,p) ,
\nn\\
\textrm{IIC}:&\quad(u,x,z)\quad\textrm{with}\quad
1-p=\eou\,z ,
\nn\\
\textrm{III}:&\quad(u,y,p)\quad\textrm{with}\quad
1-x=\eou\,y .
\end{align}
In each region, we Laurent-expand the integrand with respect to
$\veps$ by regarding the specified integration variables kept fixed.
The integration ranges, $[0,1]$ for $x$ and $p$, and $[0,u/\veps]$ for
$y$ and $z$, should be appropriately modified to avoid overlaps
among the seven regions as given in \eqref{eq:(I)}--\eqref{eq:(III)}
for $x$ and $y$.
Finally, the $u$-integration should be carried out in the range
$u>\veps$ as given there.


\begin{thebibliography}{99}


\bibitem{Schnabl}
  M.~Schnabl,
  ``Analytic solution for tachyon condensation in open string field
  theory,''
   Adv.\ Theor.\ Math.\ Phys.\  {\bf 10}, 433 (2006)  [hep-th/0511286].

\bibitem{ES09}
  T.~Erler and M.~Schnabl,
  ``A Simple Analytic Solution for Tachyon Condensation,''
   JHEP {\bf 0910}, 066 (2009)  [arXiv:0906.0979 [hep-th]].

\bibitem{Okawa}
  Y.~Okawa,
  ``Comments on Schnabl's analytic solution
    for tachyon condensation in Witten's open string field theory,''
    JHEP {\bf 0604}, 055 (2006)  [hep-th/0603159].

\bibitem{MS1}
  M.~Murata and M.~Schnabl,
  ``On Multibrane Solutions in Open String Field Theory,''
  Prog.\ Theor.\ Phys.\ Suppl.\  {\bf 188}, 50 (2011)
  [arXiv:1103.1382 [hep-th]].

\bibitem{HKwn}
  H.~Hata and T.~Kojita,
  ``Winding Number in String Field Theory,''
  JHEP {\bf 1201}, 088 (2012)  [arXiv:1111.2389 [hep-th]].

\bibitem{MS2}
  M.~Murata and M.~Schnabl,
  ``Multibrane Solutions in Open String Field Theory,''
  JHEP {\bf 1207}, 063 (2012)  [arXiv:1112.0591 [hep-th]].

\bibitem{HKinfty}
  H.~Hata and T.~Kojita,
  ``Singularities in K-space and Multi-brane Solutions in
    Cubic String Field Theory,''
  JHEP {\bf 1302}, 065 (2013)  [arXiv:1209.4406 [hep-th]].

\bibitem{EllSch}
  I.~Ellwood and M.~Schnabl,
  ``Proof of vanishing cohomology at the tachyon vacuum,''
  JHEP {\bf 0702}, 096 (2007)  [hep-th/0606142].

\bibitem{EM14}
  T.~Erler and C.~Maccaferri,
  ``String Field Theory Solution for Any Open String Background,''
  JHEP {\bf 1410}, 029 (2014)
  [arXiv:1406.3021 [hep-th]].

\bibitem{BV1}
  I.~A.~Batalin and G.~A.~Vilkovisky,
  ``Gauge Algebra and Quantization,''
  Phys.\ Lett.\ B {\bf 102}, 27 (1981).


\bibitem{BV2}
  I.~A.~Batalin and G.~A.~Vilkovisky,
  ``Quantization of Gauge Theories with Linearly Dependent
  Generators,''
  Phys.\ Rev.\ D {\bf 28}, 2567 (1983)
  [Phys.\ Rev.\ D {\bf 30}, 508 (1984)].

\bibitem{KugoOjima}
  T.~Kugo and I.~Ojima,
  ``Local Covariant Operator Formalism of Nonabelian Gauge Theories
  and Quark Confinement Problem,''
  Prog.\ Theor.\ Phys.\ Suppl.\  {\bf 66}, 1 (1979).


\bibitem{HataTera}
  H.~Hata and S.~Teraguchi,
  ``Test of the absence of kinetic terms around the tachyon vacuum in
  cubic string field theory,''
  JHEP {\bf 0105}, 045 (2001)
  [hep-th/0101162].

\bibitem{EllwoodTaylor}
  I.~Ellwood and W.~Taylor,
  ``Open string field theory without open strings,''
  Phys.\ Lett.\ B {\bf 512}, 181 (2001)
  [hep-th/0103085].

\bibitem{GiustoImbimbo}
  S.~Giusto and C.~Imbimbo,
  ``Physical states at the tachyonic vacuum of open string field theory,''
  Nucl.\ Phys.\ B {\bf 677}, 52 (2004)
  [hep-th/0309164].


\bibitem{Witten}
  E.~Witten,
  ``Noncommutative Geometry and String Field Theory,''
  Nucl.\ Phys.\ B {\bf 268}, 253 (1986).

\bibitem{Schwarz}
  A.~S.~Schwarz,
  ``Geometry of Batalin-Vilkovisky quantization,''
  Commun.\ Math.\ Phys.\  {\bf 155}, 249 (1993)
  [hep-th/9205088].


\bibitem{HataZwiebach}
  H.~Hata and B.~Zwiebach,
  ``Developing the covariant Batalin-Vilkovisky approach to string
  theory,''
  Annals Phys.\  {\bf 229}, 177 (1994)
  [hep-th/9301097].

\end{thebibliography}
\end{document}